\documentclass[aoas,MSNbibl,nameyear,seceqn,dvips]{arximspdf}
\usepackage{mathbh,multirow,dcolumn}
\usepackage{graphicx}

\doi{10.1214/13-AOAS682} %kopijuoti is PTS
\volume{7}
\issue{4}
\pubyear{2013}
\firstpage{2402}
\lastpage{2430}

\makeatletter
\newcolumntype{d}[1]{D{.}{.}{#1}}

\newcommand{\rrvert}{\vert}
\newcommand{\llvert}{\vert}
\def\cal{\mathcal}

\newcommand{\hsp}{}
\newcommand{\hspp}{}
\newcommand{\hsppp}{}
\newcommand{\Sb}{\mathbf{S}}
\newcommand{\indic}{\mathbh{1}}
\newcommand{\argmax}{\mathop{\arg\max}}
\makeatother

\begin{document}
\begin{frontmatter}

\title{Hidden Markov models for the activity profile of terrorist
groups\thanksref{T1}}
\pdftitle{Hidden Markov models for the activity profile of terrorist
groups}
\runtitle{Hidden Markov models}
\thankstext{T1}{Supported in part by the U.S. Defense Threat Reduction
Agency (DTRA) under Grant HDTRA-1-10-1-0086,
the U.S. Defense Advanced Research Projects Agency under Grant
W911NF-12-1-0034, the U.S. National Science Foundation under Grant
DMS-12-21888 and the U.S. Air Force Office of Scientific Research (AFOSR)
via the MURI grant FA9550-10-1-0569 at the University of Southern California.}

\begin{aug}
\author[a]{\fnms{Vasanthan}~\snm{Raghavan}\corref{}\thanksref{m1}\ead[label=e1]{vasanthan\_raghavan@ieee.org}},
\author[b]{\fnms{Aram}~\snm{Galstyan}\thanksref{m1}\ead[label=e2]{galstyan@isi.edu}}
\and\break
\author[c]{\fnms{Alexander G.}~\snm{Tartakovsky}\thanksref{m1,m2}\ead[label=e3]{tartakov@usc.edu}\ead[label=e4]{a.tartakov@uconn.edu}}
\runauthor{V. Raghavan, A. Galstyan and A.~G. Tartakovsky}
\affiliation{University of Southern California\thanksmark{m1}
and University of Connecticut\thanksmark{m2}}
\address[a]{V. Raghavan\\
Department of Mathematics\\
University of Southern California\\
3620 South Vermont Avenue\\
Los Angeles, California 90089 \\
USA\\
\printead{e1}}

\address[b]{A. Galstyan\\
Information Sciences Institute\\
University of Southern California\\
4676 Admiralty Way \\
Marina del Ray, California 90292 \\
USA\\
\printead{e2}}

\address[c]{A.~G. Tartakovsky\\
Department of Mathematics\\
University of Southern California\\
Los Angeles, California 90089\\
USA\\
and\\
Department of Statistics\\
University of Connecticut\\
215 Glenbrook Road\\ %U-4120 \\
Storrs, Connecticut 06071\\
USA\\
\printead{e3}\\
\phantom{E-mail:\ }\printead*{e4}}
\end{aug}

% HISTORY:
\received{\smonth{7} \syear{2012}}
\revised{\smonth{8} \syear{2013}}

% ABSTRACT
%
\begin{abstract}
The main focus of this work is on developing models for the activity
profile of a terrorist group, detecting sudden spurts and downfalls in
this profile, and, in general, tracking it over a period of time. Toward
this goal, a $d$-state hidden Markov model (HMM) that captures the latent
states underlying the dynamics of the group and thus its activity profile
is developed. The simplest setting of $d = 2$ corresponds to the case where
the dynamics are coarsely quantized as \emph{Active} and \emph{Inactive},
respectively. A state estimation strategy that exploits the underlying HMM
structure is then developed for spurt detection and tracking. This strategy
is shown to track even nonpersistent changes that last only for a short
duration at the cost of learning the underlying model. Case studies
with real
terrorism data from open-source databases are provided to illustrate the
performance of the proposed methodology.
\end{abstract}

% KEYWORDS
\end{frontmatter}

%s1 #&#
\section{Introduction}
\label{sec1}
Terrorist attacks can have an enormous impact on wide sections
of society [\citet{muellerstewart}]. Thus, the continued study of terrorism
is of utmost importance. In this direction, it is imperative to track
the activity of terrorist groups so that effective and appropriate
counter-terrorism measures can be quickly undertaken to restore order
and stability. In particular, detecting sudden spurts and downfalls in
the activity profile of terrorist groups can help in understanding
terrorist group dynamics.

The basic element toward these goals is the collation of data on
terrorist activities perpetrated by a group of interest. Over the
last decade, many databases such as \citeauthor{iterate}, the Global
Terrorism Database (GTD) [\citet{lafree1}], the RAND Database on Worldwide
Terrorism Incidents (\citeauthor{rdwti}), etc., have made data on
terrorism available in an open-source setting. However, even the best
efforts on collecting data cannot overcome issues of temporal ambiguity,
missing data and attributional ambiguity. %that leads to mislabeled
%data.
In addition, the very nature of terrorism makes it a rare occurrence from
the viewpoint of model learning and inferencing. Thus, there has been an
increased interest on parsimonious models for terrorism with strong
explanatory and predictive powers.

Prior work on modeling the activity profile of terrorist groups falls
under one
of the following three
categories. \citeauthor{enderssandler1993}
(\citeyear{enderssandler1993,enderssandler2000,enderssandler2002})
use classical time-series analysis techniques to propose a threshold
auto-regressive (TAR) model and study both the short-run as well as the
long-run spurt in world terrorist activity over the period from
1970 to 1999. On the other hand, \citet{lafreemorrisdugan}
and \citet{duganlafreepiquero} adopt group-based trajectory analysis
techniques (Cox proportional hazards model or zero-inflated Poisson
model) to
identify regional terrorism trends with similar developmental paths. The
common theme that ties both these sets of works is that the optimal
number of
underlying latent groups and the associated parameters that best fit
the data
remain variable with the parameters chosen to optimize a metric such as
the Akaike
Information Criterion (AIC), the Bayesian Information Criterion (BIC),
etc., or
via logistic regression methods. While acceptable model fits are
obtained in these
works, the complicated dependency relationships between the endogenous
and exogenous
variables makes inferencing nontransparent. In addition, both sets of
works take
a \emph{contagion} theoretic viewpoint [\citet{midlarsky,midlarsky1}]
that the current
activity of the group is \emph{explicitly} dependent on the past
history of the
group, which accounts for clustering effects in the activity profile.

The third category
that provides a theoretical foundation and an explanation for
clustering of
attacks is \citet{porterwhite2012}, where an easily decomposable two-component
self-exciting hurdle model (SEHM) for the activity profile is
introduced [\citet{hawkes1971,coxisham}]. In its simplest form, the hurdle
component of the model creates data sparsity by ensuring a prespecified density
of zero counts, while the self-exciting component induces clustering of data.
Self-exciting models have become increasingly popular in diverse fields
such as
seismology [\citeauthor{ogata} (\citeyear{ogata1,ogata})], gang behavior modeling [\citet
{mohler,cho2013}]
and insurgency dynamics [\citet{lewis}].

We propose an alternate modeling framework to the TAR model and the
SEHM by
hypothesizing that an increase (or decrease) in attack intensity can be
naturally
attributed to certain changes in the group's internal states that
reflect the
dynamics of its evolution, rather than the fact that the group has already
carried out attacks on either the previous day/week/month.
Specifically, while
the TAR model and the SEHM assume that the temporal clustering of
attacks is due
to aftershocks caused by an event, the assumption underlying our
approach is that
the clustering can be explained in terms of some unobserved dynamics of the
organization (e.g., switching to different tactics), which we address
via a
hidden Markov model (HMM) framework. We propose a $d$-state HMM for the activity
profile where each of the $d$ possible values corresponds to a certain distinct
level in the underlying attributes. The simplest nontrivial setting of
$d = 2$
with \emph{Active} and \emph{Inactive} states is shown to be a good
model that
captures most of the facets of real terrorism data.

The advantages of the proposed framework are many. The $d$-state HMM is
built on a small set of easily motivated hypotheses and is parsimoniously
described by $d(d+1)$ model parameters. While observed data rarity can be
explained by state transitions, clustering of attacks can be attributed to
different intensity profiles in the different states. In addition, the HMM
paradigm allows the use of a mature and computationally efficient toolkit
for model learning and inferencing [see, e.g., \citet{rabiner}].

Our experimental studies with two data sets (FARC and Indonesia/Timor-Leste)
suggest that, in terms of explanatory power, both HMM and SEHM perform
reasonably well, with neither framework clearly outperforming the
other. In
particular, the HMM does better for the FARC data set, whereas the SEHM
is the
better option for Timor-Leste (which is a data set where attacks are collated
across groups and thus has heavier tails in terms of severity of
attacks). On
the other hand, our results also show that the HMM approach predicts
the time
to the next day of activity more accurately than the SEHM for both data sets,
suggesting that the former model might have a better generalization capability.
While this conclusion does not imply that either framework can be
accepted (or
rejected) and a further careful study is necessary, it clearly demonstrates
that the HMM approach advocated here is a \emph{competing alternate} modeling
framework for real terrorism data.

%To address this problem, we propose two detection strategies.
%The first strategy is based on the Exponential Weighted Moving-Average
%(EWMA) filter that tracks the quantities of interest. It does not
%require
%knowledge of the underlying model and is hence useful in settings where
%model learning is difficult or where model uncertainty is high.
%However,
%as a price for this luxury, only \emph{persistent} changes can be
%detected.
%In contrast, the second strategy that
%}

In addition, we develop a strategy to quickly identify a sudden spurt (or
a sudden downfall) in the activity profile of the group. This strategy
exploits the HMM structure by learning the model parameters to estimate
the most probable state sequence using the Viterbi algorithm. We show that
this approach allows one to not only detect nonpersistent changes in
terrorist group dynamics, but also to identify key geopolitical
undercurrents that lead to sudden spurts/downfalls in a group's activity.

%f1 #&#
\begin{figure}

\includegraphics{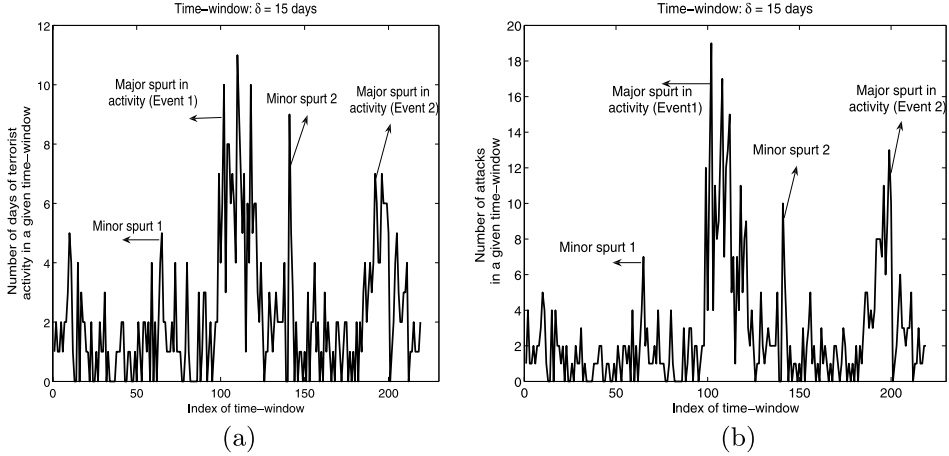}

\caption{\textup{(a)} Number of days of terrorist activity, and
\textup{(b)} Total number of attacks
in a $\delta$-day time window as a function of time for FARC ($\delta=
15$ days). The key geopolitical events in this period are also
marked.}\label{fig1}
\end{figure}

%s2 #&#
\section{Preliminaries}
\label{sec2}
%s2.1 #&#
\subsection{Qualitative features of activity profile}
\label{sec2a}

A typical example of the activity profile is presented in Figure~\ref{fig1} where
the number of days of terrorist activity in a $\delta= 15$ day time
window and
the total number of attacks within the same time window are plotted as
a function
of time. The focus of this example is \emph{Fuerzas Armadas
Revolucionarias de
Colombia} (FARC), studied %the subject of our focus
in Section~\ref{sec5a}. The data
for Figure~\ref{fig1} is obtained from the RDWTI and corresponds to incidents
over the nine-year period from 1998 to 2006. From Figure~\ref{fig1}
and a
careful study of the activity profile of many terrorist groups from similar
databases, we highlight some of the important features of terrorism
data that
impact modeling:
\begin{itemize}
\item\emph{Temporal ambiguity}: The exact instance (time) of
occurrence of
a terrorism incident is hard to pinpoint. This is because accounts
of most terrorist events are from third-party sources. Thus, the granularity
of incident reportage (i.e., the time scale on which incidents are reported)
that is most relevant is \emph{discrete}, typically days.

\item\emph{Attributional ambiguity}: Further, in many of the
databases,
there exists an ambiguity in attributing a certain terrorism incident
to a
specific group when multiple groups contest on the same geographical
turf. Some
of this ambiguity can be resolved by attributing an incident to a
specific group
based on the attack signature (attack target type, operational details,
strategies
involved, etc.). However, this is an intensive manual exercise and
there is
necessarily a certain ambiguity in this resolution.

\item\emph{Data sparsity}: Despite the recent surge in media
attention on
trans-national terrorist activities and insurgencies, terrorism
incidents are
``rare'' (from the perspective of model learning) even for some of the most
active terrorist groups. For example, the data in Figure~\ref{fig1} corresponds
to 604 incidents over a nine-year period leading to an average of
$\approx\!1.29$
incidents per week. While a case can be made that these databases
report only a
subset of the true activity, the fact that significant amount of
resources have
to be invested by the terrorist group for every new incident acts as a natural
dampener toward more attacks.
\end{itemize}
These three features make a strong case for parsimonious models for the activity
profile of terrorist groups. Further, any good model should be robust
to a small
number of errors in terms of mislabeled data and missing data supplied
from other
terrorism incident databases.

%s2.2 #&#
\subsection{Prior work}
\label{sec2b}
%Aram
%The typical observable of a terrorist group is its activity profile
%and the previous discussion showed why a model for terrorism
%cannot be developed in continuous-time.
%A point process model in
%discrete-time \citep{cox_isham} is thus the most general framework to
%model the activity profile.
The activity profile of a terrorist group can be modeled as a
discrete-time stochastic process.
Let the first and last day of the time period of interest be denoted
as Day $1$ and Day ${\cal N}$, respectively. Let $M_i$ denote the number
of terrorism incidents on the $i$th day of observation, $i = 1, \ldots
, {\cal N}$.
Note that $M_i$ can take values from the set $\{ 0, 1, 2, \ldots\}$ with
$M_i = 0$ corresponding to no terrorist activity on the $i$th day of
observation. On the other hand, there could be multiple terrorism incidents
corresponding to independent attacks on a given day, reflecting a
high level of coordination between various subgroups of the group. Let
${\cal H}_i$ denote the history of the group's activity till day $i$.
That is,
${\cal H}_i =  \{M_1, \ldots, M_i  \}, \hsppp i = 1,2,
\ldots, {\cal N}$
with ${\cal H}_0 = \varnothing$. The point process model is complete if
$\mathsf{ P} (M_i = r | {\cal H}_{i-1}  )$ is specified as a
function of
${\cal H}_{i - 1}$ for all $i = 1, \ldots, {\cal N}$ and $r = 0,1,2,
\ldots.$

We noted in Section~\ref{sec1} that prior work on models for the
activity profile
fall under one of three categories. In the time-series techniques pioneered
by \citet{enderssandler2000}, a nonlinear trend component, a seasonality
(cyclic) component and a stationary component are fitted to the time-series
data of worldwide terrorism incidents. In particular, the following model-fit
is proposed for $\{ M_i \}$:
\[
\sum_{ i \hsppp\in\Delta_q } M_i = \sum
_{j = 0}^n \alpha_j q^j +
\beta\sin ( \omega q + \phi ) + \mu_q,
\]
where $\Delta_q$ denotes the period corresponding to the $q$th quarter
in the
period of observation and $ \{ n, \hsppp\alpha_0, \hsppp\ldots
, \hsppp
\alpha_n, \hsppp\beta, \hsppp\omega, \hsppp\phi, \hsppp\mathsf{
Var}(\mu_q)
\}$ are parameters to be optimized over some parameter
space. This modeling effort results in an eight-parameter model (4 for the
trend component, 3 for the seasonality component and a variance
parameter for
the stationary component) which is then used to identify a rough $4$ and
$1/2$-year cycle in terrorism events. Further, a nonlinear trend and
seasonality component ensures that trends in terrorism cannot be predicted,
thus explaining the observed boom and bust cycles in terrorist activity.
Alternately, a TAR model that switches from one auto-regressive process to
another with the switches corresponding to key geopolitical events is studied
in \citet{enderssandler2002}.

%In \citep{dugan_lafree_piquero}, the hazard function $\mathsf{ Haz}(t)$ of
%the time
%to the next hijacking (denoted as the random variable $T$) is modeled
%as
%{\rm Context},
%( t \leq T < t + \Delta t) } {\Delta t},
%where $\mathsf{ Haz}_0$ is a baseline hazard function, and $\{ \beta_{i}
%are optimized from a certain model class.}

%Aram
On the other hand, the SEHM of \citet{porterwhite2012} is described as
%
%
%e2.1 #&#
\begin{equation}
\mathsf{ P} ( M_i = r | {\cal H}_{i-1} ) = \cases{ %
e^{- (B_i + \operatorname{SE}_i({\cal H}_{i-1}))}, &\quad $r = 0,$
\vspace*{2pt}\cr
\displaystyle\frac{r^{-s}}{\zeta(s)} \cdot \bigl(1 - e^{- (B_i + \operatorname{SE}_i({\cal
H}_{i-1}) )} \bigr), &\quad $r \geq1,$} %\nonumber
\label{sehmeqn}
\end{equation}
where $B_i$ is a baseline process, and $\operatorname{SE}_i(\cdot)$ is the
self-exciting component
given as
\[
\operatorname{SE}_i ( {\cal H}_{i-1} ) = \sum
_{ j \hsppp: \hsppp j \hsppp< \hsppp i,
\hsppp M_j \hsppp> \hsppp0} \alpha_j g (i-j)
\]
for an appropriate choice of decay function $g(\cdot)$ and influence
parameters $\{ \alpha_j \}$. %Note that the Zipf distribution falls
%in the class of power-law %(heavy tailed)
%distributions (see Supplementary
%Part A) and is also explored in \citep{clauset}.
On the other hand, $s \in(1,\infty)$ is an appropriately chosen
parameter of
the %Zipf\footnote{This distribution is sometimes referred to as
%discrete Pareto.}
zeta distribution, and $\zeta(s) = \sum_{n = 1}^{\infty} n^{-s}$ is the
Riemann-zeta function. While a constant $s$ parameter leads to the simplest
modeling framework, $s$ can in general be driven by another self-exciting
process. %(see \citet{porter_white2012} for details).
A class described by eight parameters is studied in \citet{porterwhite2012}
and it is shown that a four-parameter model optimizes an AIC metric for
terrorism data from Indonesia/Timor-Leste over the period from 1994 to 2007.
This model is shown to accurately capture terrorism data (especially the
extreme outliers such as days with $36, 11$ and $10$ attacks). The heavy-tailed
zeta distribution is also explored in \citet{clauset} for modeling extremal
terrorist events.

\section{Proposed model for the activity profile}
\label{sec3}
While the above set of models capture terrorism data, we now propose a
\emph{competing alternate} framework based on HMMs. Our proposed
framework is based on the following simplifying
hypotheses:
\begin{itemize}
\item\emph{Hypothesis} 1: The activity profile of a terrorist group
$\{ M_i, \hsppp i = 1, \ldots, {\cal N} \}$ depends only on certain
states ($\Sb_i$) in the sense that the current activity is
conditionally independent of the past activity/history of the
group given the current state:
\[
\mathsf{ P} (M_i | {\cal H}_{i-1}, \hsppp\Sb_i )
= \mathsf{ P} (M_i | \Sb_i ), \qquad i = 1,2, \ldots.
\]
In other words, these states completely capture the facets from the
past history of the group in determining the current state of activity.
While attaching specific meanings to $\Sb_i$ is not the focus of this
paper, an example in this direction is the postulate by \citet{rand}
that the \emph{Intentions} and the \emph{Capabilities} of a group could
serve as these states.

\item\emph{Hypothesis} 2: The dynamics of terrorism are
well understood if the underlying states $\{ \Sb_i \}$ are known. However,
in reality, $\Sb_i$ cannot be observed directly and we can only make
indirect inferences about it by observing $\{ M_j, j = 1, \ldots, i \}$.
To allow inferencing of $\Sb_{i}$, we propose a $d$-state model
that captures the dynamics of the group's latent states over time. That
is, $\Sb_{i} \in\{ 0, 1, \ldots, d-1 \}$ with each distinct value
corresponding to a different level in the underlying attributes of the
group. Further, we penalize frequent state transitions in the terrorist
group dynamics by constraining $\Sb_{i}$ to be fixed over a block
(time window) of $\delta$ days, where $\delta$ is chosen appropriately
based on the group dynamics.
\end{itemize}

%{}
%
%f2 #&#
\begin{figure}

\includegraphics{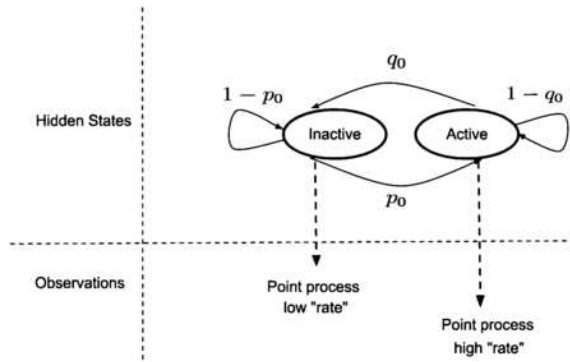}

\caption{A simple $d = 2$ state HMM for the activity of a terrorist
group.}\label{fighmm}
%A simple $d = 2$ state hidden Markov model for the \emph{Capabilities}
%$S_{1,i}$ of a terrorist group.
\end{figure}

A typical illustration of this framework with $d = 2$ is provided in
Figure~\ref{fighmm}, where the state over the $n$th time window
($\Delta_n, \hspp n = 1, 2, \ldots, K$) corresponding to $\Delta_n =
\{ (n-1)\delta+1, \ldots, n\delta\}$ and $K = \lceil\frac{ {\cal N}} {
\delta}\rceil$ is given as
\[
\Sb_{ i} |_{ i \in\Delta_n } = s_n,\qquad  s_n \in
\{ 0, 1 \}.
\]
% = (n-1)\delta+ 1, \cdots, n\delta, \hspp n = 1, 2, \cdots.
%
In the \emph{Inactive} state ($s_n = 0$), the underlying $M_i$ form a
low-``rate'' point process, whereas in the \emph{Active} state ($s_n = 1$),
the $M_i$ form a high-``rate'' point process. Thus, a state transition
from $s_{n-1} = 0$ to $s_n = 1$ indicates a spurt in the activity of
the group, whereas an opposite change indicates a downfall in the activity.
This evolution in the states of the group is modeled by a state
transition probability matrix $\mathbf{P} = \{ \mathsf{ P}_{ij} \}$, where
\[
\mathbf{P} = \left[ %
\matrix{ 1 - p_0 &
p_0
\vspace*{2pt}\cr
q_0 & 1 -q_0 }
\right]
\]
with
$\mathsf{ P}_{ij} \triangleq\mathsf{ P} (s_n = j | s_{n-1} = i )
%| S_{1, \hsppp(n-1)\delta+1 } = i)
, \hspp i,j \in\{ 0,1 \}$.
In general, there exists a trade-off between accurate modeling of the
group's latent attributes (larger $d$ %and more observation density
%parameters
is better for this task) versus estimating more model
parameters (smaller $d$ is better). While %almost all the
attention in the sequel will be restricted to the $d = 2$ setting because
of the implementation ease of the proposed approaches in this setting,
these approaches can be easily extended to a general $d$-state setting.

To model the observations in either state [i.e., $\mathsf{ P}(M_i |
\Sb_{i} = j), \hspp j = 0,1$], a geometric density can be motivated with
the following hypothetical scenario. Consider a setting where the group
has infinite resources\setcounter{footnote}{1}\footnote{While the infinite resource assumption
is impractical in capturing the dynamics of terrorist groups, it allows
us to mathematically motivate the geometric model.} and orchestrates $M_i$
attacks on the $i$th day till the success of a certain short-term policy
objective can be declared. Every additional attack contributes equally to
the success of this objective and, as long as
the group's objective has not been met, attaining this objective in the future
is not dependent on the past attacks. In other words, the group remains
\emph{memoryless} (or is \emph{oblivious}) of its past activity and continues
to attack with the same pattern as long as its objective remains unmet. A
slight modification in the group dynamics that assumes that a group
resistance/hurdle needs to be overcome before this \emph{modus operandi}
kicks in leads to a hurdle-based geometric model for~$\{ M_i \}$.

While these strategies are theoretically motivated and at best may describe
a specific group, other groups could adopt different strategies. In fact,
strategies could also change with time. Thus, a certain model could make
more sense for a given terrorist group than other models. In our subsequent
study, we consider the following one-parameter models with support on the
nonnegative integers: Poisson, shifted zeta, and geometric. We also consider
the following two-parameter models: P\`{o}lya,\footnote{Also referred
to as negative binomial with positive real parameter.} (nonself-exciting)
hurdle-based zeta, and hurdle-based geometric. Of these six models, the
geometric and the hurdle-based geometric allow simple recursions for
estimates of model parameter(s) via the Baum--Welch algorithm, while the
shifted zeta and the hurdle-based zeta distributions allow heavy tails;
see supplementary material, Part A [\citet{vasanthsuppA}]. Further details on these
models, such as the associated probability density function, log-likelihood
function, Maximum Likelihood (ML) estimate of the model parameters and a
formula for the AIC, are also provided in \citet{vasanthsuppA}. The study
of fits of these models to specific terrorist groups is undertaken in
Section~\ref{sec6b}.

%s4 #&#
\section{Detecting spurts and downfalls in activity profile}
\label{sec4}
We are interested in solving the %\emph{noncausal}
inference problem:
%
%
%e4.1 #&#
\begin{equation}
%%\{ 0, \hsppp\cdots, \hsppp d-1 \}
%%\nonumber
%}
\widehat{s}_n |_{ n = 1, \ldots, K} =
\argmax _{ s_1, \ldots, s_K \hsppp
\in\hsppp
\{0, \hsppp1 \}^K } \mathsf{ P} \bigl( \Sb_{i} = s_n,
\hsppp i \in\Delta_n | \{ M_j, \hsppp j = 1, \hsppp\cdots,
\hsppp{\cal N} \} \bigr). %\nonumber
\label{eqn2}
\end{equation}
In particular, we are interested in identifying state transitions that
correspond to either a spurt or a downfall in activity.

Since we are interested in tracking changes in the latent attributes of
the group, we focus on an observation sequence that captures the
\emph{resilience} of the group and another that reflects the \emph{level
of coordination} in the group [\citet{santos,lindberg}]. In particular,
the ability of a group to sustain terrorist activity over a number of days
reflects its capacity to rejuvenate itself from asset (manpower, material
and skill-set) losses. And the ability of the group to launch
multiple attacks over a given time period reflects its capacity to coordinate
these assets necessary for simultaneous action often over a wide geography.
That is, mathematically speaking, the focus is on: (i) $X_n$, the
number of
days of terrorist activity, and (ii) $Y_n$, the total number of
attacks, both
within the $\delta$-day time window $\Delta_n$:
\[
X_n = \sum_{i \in\Delta_n} \indic \bigl( \{
M_i > 0\} \bigr);\qquad Y_n = \sum
_{ i \in\Delta_n } M_i,\qquad n = 1,2, \ldots,
\]
where $\indic(\cdot)$ denotes the indicator
function of the set under consideration. Note that $Y_n/\delta$ is the average
number of attacks per day and, thus, $Y_n$ is a reflection of the intensity
of attacks launched by the group. In general, $X_n$ is more indicative
of resilience in the group, whereas $Y_n$ captures the level of coordination
better.

To build a model-driven detection strategy, we now develop a
probabilistic model for $X_n$ and $Y_n$. For this, we leverage the
rare nature of terrorism to hypothesize that most terrorist groups tend
to be in an \emph{Inactive} state for far longer than in an \emph{Active}
state. Thus, it is reasonable\footnote{Note that in (\ref{eqn3}) we have
not made any specific assumptions on the distribution of $M_i$. In fact,
we have only labeled the quantity in (\ref{eqn3}) as $\gamma_j$.} to assume
that $\Sb_{i} = 0$ for long stretches of time and $\gamma_0 \approx
0$, where
%
%
%e4.2 #&#
\begin{equation}
\gamma_j \triangleq \mathsf{ P}(M_i > 0 |
\Sb_{i} = j ), \qquad j = 0,1. \label{eqn3} %\nonumber
\end{equation}
Over such a long stretch where $\Sb_{ i} = 0$, an elementary consequence
of (\ref{eqn3}) is that $X_n$ is a binomial random variable with
parameters $\delta$ and $\gamma_0$:
%(that is, $X_n \sim{\rm Binomial}(\delta, \hsppp\gamma_0)$).
%
\[
\mathsf{ P}(X_n = k) = \pmatrix{\delta\cr k} \cdot(\gamma_0
)^k \cdot ( 1 - \gamma_0 )^{\delta- k}.
\]
If $\delta$ is sufficiently large (typical values used in subsequent
case studies are $\delta= 15$ to $30$ days) so that the binning/edge
effects can be neglected, $X_n$ can be well approximated by a Poisson
random variable with rate parameter $\delta\gamma_0$. In fact, we have
the following bound [\citet{teera}, Corollary 3.2] on the
approximability of
the binomial distribution by Poisson for $k = 0, \ldots, \delta$:
%
%
%e4.3 #&#
\begin{equation}
\bigl|\mathsf{ P}_\mathsf{ Bin}(X_n = k) - \mathsf{ P}_\mathsf{ Poi}(X_n
= k)\bigr| \leq\min \biggl(1 - e^{-\delta\gamma_0}, \hsppp\frac{\delta\gamma_0} {
k} \biggr) \cdot
\gamma_0 \approx0. %\nonumber
\label{poibinomial}
\end{equation}
Equivalently, let $T_k, \hsppp k = 1,2, \ldots $ denote the time to the
$k$th day of terrorist activity (with $T_0$ set to $T_0 = 0$). Then,
$\Delta T_k = T_k - T_{k-1}$ denotes the time %number of days of wait
%to the
to the subsequent day of activity (inter-arrival duration) and
$\Delta T_k$ is approximately exponential with mean $1/\gamma_0$.
While a similar reasoning suggests that $\Delta T_k$ in the \emph{Active}
state is exponential with mean $1/\gamma_1$, this fit is bound to be good
only in the first-order sense because a terrorist group is expected to
stay in the \emph{Active} state for relatively shorter durations and
$\gamma_1 \gg \gamma_0$. Rephrasing the above conclusions, a discrete-time
Poisson process model is a good model for the days of terrorist activity,
especially in the \emph{Inactive} state.

Under the same assumptions (as above), in the \emph{Inactive} state, $Y_n$
can be rewritten as
\[
Y_n = \sum_{ i \hsppp\in\hsppp{\cal A}_n } M_i,
\]
where ${\cal A}_n \subset\Delta_n$ is the set of days of activity in
$\Delta_n$ with $| {\cal A}_n | = X_n$. Thus, $Y_n$ can be approximated
as a compound Poisson random variable [\citet{coxisham}] whose density
is expressed in terms of the density function of $M_i$. For example, if
$M_i$ is independent and identically distributed (i.i.d.) as geometric
with $\mathsf{ P} ( M_i = k |\Sb_{i} = 0  )
= (1 - \gamma_0) \cdot(\gamma_0)^k, \hspp k \geq0$, a simple
computation [see \citet{vasanthsuppA}] shows that
\[
\mathsf{ P}(Y_n = r)  =  (1 - \gamma_0)^{\delta}
\cdot ( \gamma _0 )^r \cdot\pmatrix{\delta+ r - 1 \cr r},\qquad
\hspp r \geq0.
\]
Similarly, the joint density of $(X_n, \hsppp Y_n)$ can be written as
\[
\mathsf{ P}( X_n = k, \hsppp Y_n = r )  =  (1 -
\gamma_0)^{\delta} (\gamma_0)^r \cdot\pmatrix{
\delta\cr k} \cdot\pmatrix{ r - 1 \cr r - k}, \qquad r \geq k,
\nonumber
\]
where the condition $r \geq k$ ensures that at least one attack occurs
on a day of activity. With the more general hurdle-based geometric model,
where
\[
\mathsf{ P}(M_i = k | \Sb_i = 0) = \cases{ %
 1 - \gamma_0, & \quad $k = 0,$
\vspace*{2pt}\cr
\gamma_0 (1 - \mu_0) \cdot(\mu_0)^{k-1},
&\quad  $k \geq1,$ }
\]
the joint density is given as
\begin{eqnarray}
\mathsf{ P}( X_n = k, \hsppp Y_n = r ) = \pmatrix{\delta\cr k} \pmatrix{
r -1 \cr r - k} \cdot(1 - \gamma_0)^{\delta - k} (
\gamma_0)^{k} \cdot(1 - \mu_0)^k(\mu_0)^{r - k},\nonumber\\
 \eqntext{r \geq k.}
\end{eqnarray}
Replacement of $\gamma_0$ and $\mu_0$ with $\gamma_1$ and $\mu_1$
in the
\emph{Active} state works subject to the same issues/constraints as
stated earlier.

We now propose a strategy that exploits the underlying HMM structure to detect
changes in group dynamics. For this, we treat as observations $\{ X_n \}$,
$\{ Y_n \}$ and the joint sequence $\{ (X_n, \hsppp Y_n) \}$ under different
modeling assumptions on $\{ M_i \}$. We first apply the classical HMM
formulation [\citet{rabiner}] where the Baum--Welch algorithm is used
to learn the
parameters that determine the density function of the observations. For the
Baum--Welch algorithm to converge to a local maximum (with respect to the
log-likelihood function) in the parameter space, a sufficient condition is
that the density function of the observation be
log-concave [\citet{rabiner}, Section IVA, page 267]. %In Supplementary
%Part A, u
Under the i.i.d. geometric and hurdle-based geometric models for $\{
M_i \}$,
it is established in \citet{vasanthsuppA} that all the three density functions
are log-concave. Further, an iterative update for
the parameter estimates is also established under these two models.
With the
converged Baum--Welch parameter estimates as the initialization, the Viterbi
algorithm is then used to estimate the most probable state sequence
given the
observations. The output of the Viterbi algorithm is a state estimate
for the
period of interest
\[
\bigl\{ \Sb_i = \widehat{s}_n \hsppp\in\hsppp\{ 0,
\hsppp1 \}\ \mathrm{for}\ \mathrm{all}\ \hspp i \in\Delta_n\ \mathrm{and}\ \hspp n = 1, \ldots, K \bigr\}.
\]
A state estimate of $1$ indicates that the group is \emph{Active} over the
period of interest, whereas an estimate of $0$ indicates that the group is
\emph{Inactive}. Transition between states indicates spurt/downfall in
the activity.

While we have developed a model-driven strategy, there is often an interest
in approaches that are independent of these parameters, and hence not sensitive
to the parameter estimation algorithms or the length of the training period.
An alternate approach that does not explicitly learn the underlying model
parameters can be developed using the Exponential Weighted Moving-Average
(EWMA) filter. Details of this approach and its comparative performance with
the HMM-based strategy can be found in the work of \citet{vasantharxiv2012}.

%s5 #&#
\section{Case-studies}
\label{sec5}
We now undertake two case studies on the fit of a discrete-time
Poisson point process model for the days of terrorist activity. For
this, we classify terrorist activities as reported in the RDWTI and GTD
that catalogue activities by different groups across the
world [\citet{lafree1}, \citeauthor{rdwti}].

%s5.1 #&#
\subsection{FARC}
\label{sec5a}
We start with FARC, a Marxist-Leninist terrorist group active over a
large area
in Colombia and its neighborhood. We study the activities of FARC over the
nine-year time period from 1998 to 2006. This period covers a total of 604
terrorist incidents in the RDWTI with a yearly breakdown of 44 incidents
in 1998, 18 in 1999, 45 in 2000, 27 in 2001, 217 in 2002, 57 in 2003,
33 in 2004, 66 in 2005 and 97 in 2006, respectively. The reasons why FARC
and the 1998 to 2006 time period have been chosen for our study are provided
in supplementary material, Part B [\citet{vasanthsuppB}].

%f3 #&#
\begin{figure}

\includegraphics{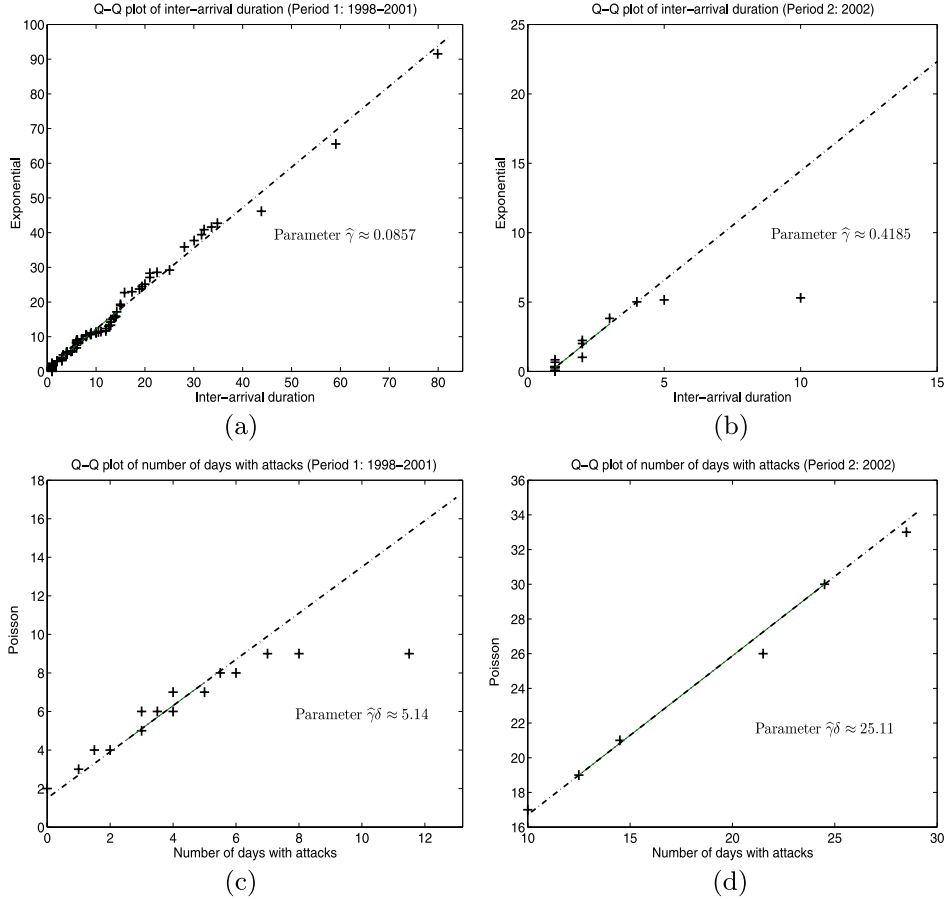}

\caption{Q--Q plots of %inter-arrival
duration between days of terrorist activity with respect to a theoretical
exponential random variable for the period from: \textup{(a)} 1998 to 2001
corresponding to normal terrorist activity, and \textup{(b)} 2002 corresponding
to a spurt in terrorist activity. Q--Q plots of number of days of terrorist
activity over a $\delta= 60$ day time window with respect to a theoretical
Poisson random variable for the same two scenarios are presented in
\textup{(c)} and~\textup{(d)}.}
\label{fig2}
\end{figure}
%
%%\begin{minipage}{5.5in}
%%\centerline{
%&
%%}
%%\end{minipage}
%%\begin{minipage}{5.5in}
%%\centerline{
%&
%%}
%%\end{minipage}
%
%
As explained in Section~\ref{sec4}, the activity profile of FARC can be
modeled as a discrete-time sampled Poisson point process. In
Figure~\ref{fig2}, we test the fit of this model by studying: (i) the
Quantile--Quantile (Q--Q) plots comparing the sample inter-arrival
duration between consecutive days of terrorist activity and an
exponential random variable with an appropriately-defined rate
parameter ($\widehat{\gamma}$), and (ii)~the Q--Q plots comparing the number
of days of terrorist activity over a time window of $\delta$ days and
a Poisson random variable with parameter $\delta\widehat{\gamma}$. For
both sets of Q--Q plots, we consider two scenarios: the first (the
1998 to 2001 period) corresponding to a period of ``normal'' terrorist
activity, and the second (the 2002 period) corresponding to a spurt in
terrorist activity.

%%\begin{minipage}{5.5in}
%%\centerline{
%&
%%\\
%%\includegraphics[height=2.2in,width=2.6in]
%{interarrival_farc_states_classified.eps}
%%&
%%\includegraphics[height=2.2in,width=2.6in]
%{no_attacks_farc_states1mod.eps}
%%\\
%%(c) & (d)
%%}
%%\end{minipage}
%(a) Tracking properties of $R_{2,n}$ in response to changes in $Y_n$
%as a function of the smoothing parameter, (b) Performance of the
%three EWMA tests in spurt detection for FARC data ($\delta=
%15$ days is used).
%%(c) Inter-arrival duration and state classification
%%via Viterbi algorithm, (d) Number of days of terrorist activity in a
%%$\delta$-day time-window and state re-classification.
%}
%
%}

Figures~\ref{fig2}(a) and (b) compare the Q--Q plots of the sample
inter-arrival duration under these two scenarios with an exponential
random variable. The rate parameter used for the exponential
%random variable
is
%
%
%e5.1 #&#
\begin{equation}
\label{est1} %\widehat{\rho}
\widehat{\gamma} = \frac{1}{ \mathsf{ E}[ \Delta T_k]},
\end{equation}
where $\mathsf{ E}[\Delta T_k]$ is the sample mean of the inter-arrival duration
over the considered period. From Figure~\ref{fig2}, we note that both in
periods of normal as well as a spurt in activity, the Q--Q plot
is nearly linear with a few sample outliers in the tails. These outliers
indicate that an exponential model for inter-arrival duration is not
accurate because of the heavy tails. Nevertheless, to a first-order, an
exponential random variable serves as a good fit for the inter-arrival
durations. Our numerical study leads to the following estimates for
$\gamma_{\bullet}$: (a) $\widehat{\gamma} \approx0.0857$,
(b) $\widehat{\gamma} \approx0.4185$, suggesting that a spurt in
activity is associated with an increase
in the rate parameter. Similarly, in Figures~\ref{fig2}(c) and (d), we
compare the Q--Q plots of the number of days of terrorist activity over a
$\delta= 60$ day time window under the same scenarios (as above) with a
theoretical Poisson random variable of parameter $\delta\hsppp
\widehat
{\gamma} = 60 \hsppp\widehat{\gamma}$.
While we observe some outliers in the tail quantiles, a Poisson random
variable seems to be a good first-order fit for the number of days of
terrorist activity.\looseness=-1

As explained in Section~\ref{sec4}, a geometric model is assumed for
$\{ M_i \}$ and the Baum--Welch algorithm is used to learn the underlying
$\gamma_j$ with $\{ X_n \}$, $\{ Y_n \}$ and $\{ (X_n, \hsppp Y_n) \}$
over a given $\delta$-day time window as training data. Specifically,
Table~\ref{table01} summarizes the parameter estimates for different
$\delta$ values when $\{ (X_n, \hsppp Y_n ) \}$ is used as a training set.
It is to be noted that the learned parameter values remain stable
across a large range of $\delta$ ($1$ to $30$ days) values. The
performance of the Baum--Welch algorithm is also robust to both the
length of the training set as well as the initialization. Further,
the parameter values also remain essentially independent of whether
$\{ X_n \}$, $\{ Y_n \}$ or $\{ (X_n, \hsppp Y_n ) \}$ is used for
training. For example, with $\delta= 15$ and $\{ (X_n, \hsppp Y_n ) \}$
as training data,
the parameter values learned are $\widehat{\gamma}_0 = 0.0924$ and
$\widehat{\gamma}_1 = 0.3605$, whereas with $\{ X_n \}$, these values
are $\widehat{\gamma} _0 = 0.0941$ and $\widehat{\gamma}_1 = 0.3861$.
This observation is not entirely surprising since $\{ M_i \}$ is assumed
to come from a one-parameter model family and conditioned on one of the
two variables ($X_n$ or $Y_n$), the other variable adds no significant
new information about the model parameter.

%t1 #&#
\begin{table}
\tabcolsep=3pt
\caption{State classification of FARC with the geometric and hurdle-based
geometric models and $\{ (X_n, \hsppp Y_n) \}$ as observations for
different time window periods ($\delta$)}
\label{table01}
%{}
%
%
\begin{tabular*}{\textwidth}{@{\extracolsep{\fill}}lccccc@{}}
\hline
& \multicolumn{2}{c}{\textbf{Parameters learned}} &&&\\[-6pt]
& \multicolumn{2}{c}{\hrulefill} &&&\\[-8pt]
\multicolumn{1}{@{}l}{\multirow{2}{25pt}{$\bolds{\delta}$ \textbf{(in days)}}} &
\multicolumn{1}{c}{\multirow{1}{25pt}[-10pt]{\centering{$\bolds{\widehat{\gamma}_0}$}}}
& \multicolumn{1}{c}{\multirow{1}{25pt}[-10pt]{\centering{$\bolds{\widehat{\gamma}_1}$}}} &
\multicolumn{1}{c}{\multirow{3}{75pt}[10pt]{\centering{\textbf{Length of training set (}$\bolds{N}$~\textbf{time
windows)}}} }
&
\multicolumn{1}{c}{\multirow{3}{60pt}[10pt]{\centering\textbf{No. of \emph{Active}  states (}$\bolds{\mathsf{ N}_\mathsf{ spurt}}$
\textbf{time windows)}}}&
\multicolumn{1}{c@{}}{\multirow{3}{50pt}[10pt]{\centering\textbf{Fractional activity} $\bolds{(\mathsf{ f} =
\frac{ \mathsf{ N}_\mathsf{ spurt} \cdot\delta}{ {\cal N}})}$}}\\\\
\hline
\phantom{0}$1$ & $0.0924$ & $0.3597$ & $3286$ & $483$ & $0.1469$ \\ %\hline
\phantom{0}$5$ & $0.0919$ & $0.3584$ & \phantom{0}$657$ & $140$ &$0.2130$ \\ %\hline
\phantom{0}$7$ & $0.0924$ & $0.3598$ & \phantom{0}$469$ & \phantom{0}$89$ & $0.1895$ \\ %\hline
$10$ & $0.0911$ & $0.3568$ & \phantom{0}$328$ & \phantom{0}$73$ & $0.2221$ \\ %\hline
$15$ & $0.0924$ & $0.3605$ & \phantom{0}$219$ & \phantom{0}$46$ & $0.2099$ \\ %\hline
$25$ & $0.0908$ & $0.3592$ & \phantom{0}$131$ & \phantom{0}$26$ & $0.1977$ \\ %\hline
$30$ & $0.0930$ & $0.3593$ & \phantom{0}$109$ & \phantom{0}$20$ & $0.1825$ \\
\hline
\end{tabular*}
%
%
%model
%and $\{ (X_n, \hsppp Y_n) \}$ as observations for different time-window
%periods ($\delta$)}
%{}
%
%
\tabcolsep=3pt
\begin{tabular*}{\textwidth}{@{\extracolsep{4in minus 4in}}lccccccc@{}}
\hline
& \multicolumn{4}{c}{\textbf{Parameters learned}} &&&\\[-6pt]
& \multicolumn{4}{c}{\hrulefill} &&&\\[-8pt]
\multicolumn{1}{c}{\multirow{2}{25pt}{$\bolds{\delta}$ \textbf{(in days)}}} &
\multicolumn{1}{c}{\multirow{1}{25pt}[-10pt]{\centering{$\bolds{\widehat{\gamma}_0}$}}} &
\multicolumn{1}{c}{\multirow{1}{25pt}[-10pt]{\centering{$\bolds{\widehat{\mu}_0}$}}} &
\multicolumn{1}{c}{\multirow{1}{25pt}[-10pt]{\centering{$\bolds{\widehat{\gamma}_1}$}}}&
\multicolumn{1}{c}{\multirow{1}{25pt}[-10pt]{\centering{$\bolds{\widehat{\mu}_1}$}}}&
\multicolumn{1}{c}{\multirow{3}{75pt}[10pt]{\centering{\textbf{Length of training set (}$\bolds{N}$ \textbf{time windows)}}}}
&
\multicolumn{1}{c}{\multirow{3}{60pt}[10pt]{\centering\textbf{No. of \emph{Active}  states (}$\bolds{\mathsf{ N}_\mathsf{ spurt}}$ \textbf{time windows)}}}&
\multicolumn{1}{c@{}}{\multirow{3}{50pt}[10pt]{\centering\textbf{Fractional activity}
$\bolds{(\mathsf{ f} = \frac{ \mathsf{ N}_\mathsf{ spurt} \cdot\delta}{ {\cal N}})}$}}
\\\\
\hline
\phantom{0}$1$ & $0.0950$ & $0.0752$ & $0.3982$ & $0.3083$ &
$3286$ & $484$ & $0.1472$ \\ %\hline
\phantom{0}$5$ & $0.0942$ & $0.0730$ & $0.3934$ & $0.3066$ &
\phantom{0}$657$ & $140$ &$0.2130$ \\ %\hline
\phantom{0}$7$ & $0.0949$ & $0.0742$ & $0.3956$ & $0.3081$ &
\phantom{0}$469$ & \phantom{0}$87$ & $0.1853$ \\ %\hline
$10$ & $0.0936$ & $0.0724$ & $0.3891$ & $0.3044$ &
\phantom{0}$328$ & \phantom{0}$71$ & $0.2160$ \\ %\hline
$15$ & $0.0958$ & $0.0759$ & $0.4019$ & $0.3103$ &
\phantom{0}$219$ & \phantom{0}$42$ & $0.1917$ \\ %\hline
$25$ & $0.0934$ & $0.0738$ & $0.3957$ & $0.3046$ &
\phantom{0}$131$ & \phantom{0}$26$ & $0.1977$ \\ %\hline
$30$ & $0.0954$ & $0.0794$ & $0.3966$ & $0.3073$ &
\phantom{0}$109$ & \phantom{0}$20$ & $0.1825$ \\
\hline
\end{tabular*}
\end{table}

%f4 #&#
\begin{figure}

\includegraphics{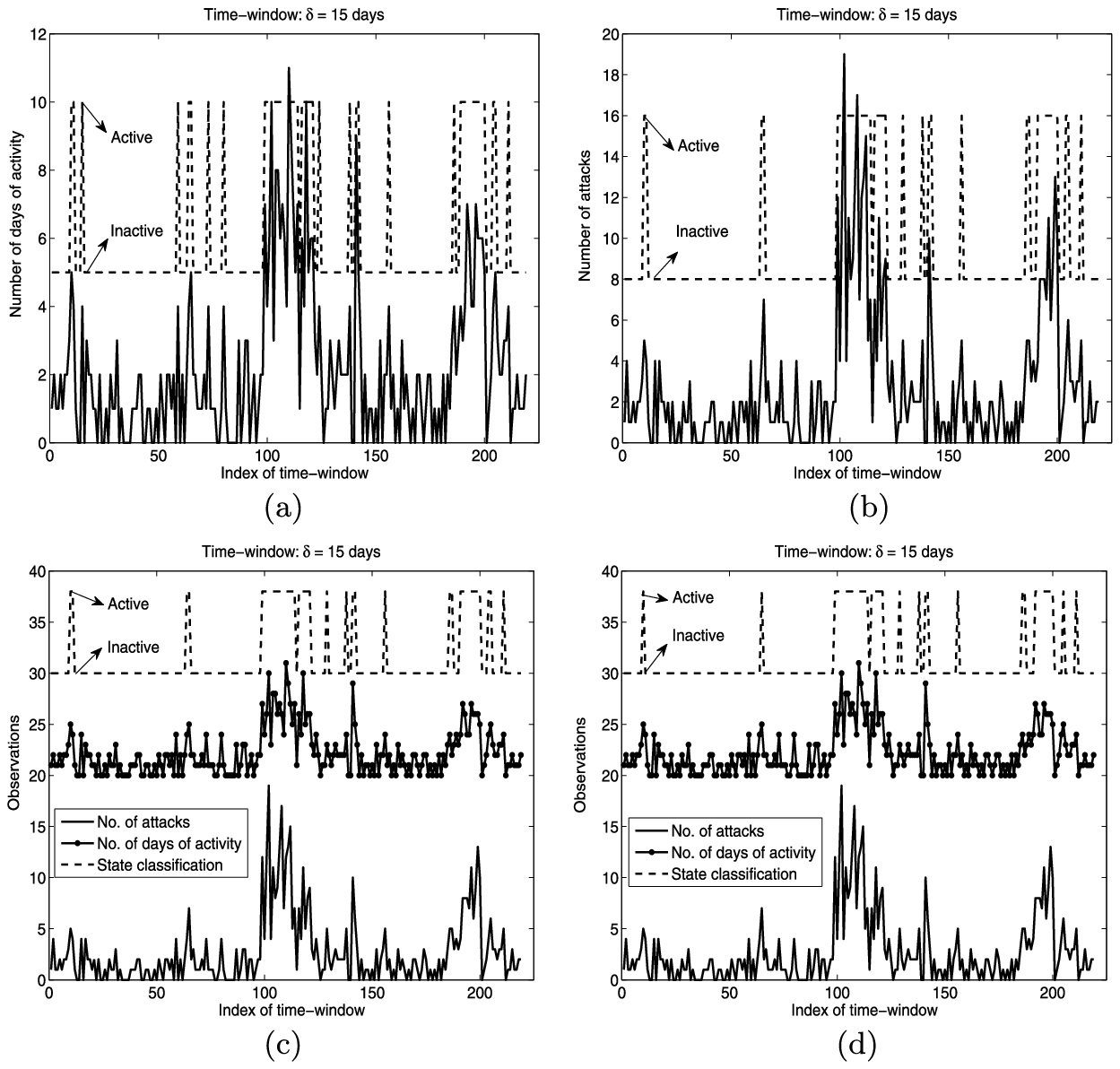}

\caption{State classification via Viterbi algorithm with the geometric model
for $\{ M_i \}$; Observations being \textup{(a)} $\{ X_n \}$, \textup{(b)} $\{ Y_n \}$
and \textup{(c)} $\{ (X_n, \hsppp Y_n ) \}$ with parameters learned by
Baum--Welch algorithm. \textup{(d)}~State classification with the hurdle-based
geometric model for the observation sequence $ \{ (X_n, \hsppp Y_n) \}
$.}\label{fig02}\vspace*{-3pt}
\end{figure}

%%\begin{minipage}{5.5in}
%%\centerline{
%{state_class_Xn_edit4_bw.eps}
%&
%{state_class_Yn_edit3_bw.eps}
%%}
%%\end{minipage}
%%\begin{minipage}{5.5in}
%%\centerline{
%{state_class_XnYn_edit4_bw.eps}
%&
%{state_class_XnYn_hbg_edit3_bw.eps}
%%}
%%\end{minipage}
%
%

The converged Baum--Welch parameter estimates are then used to initialize
the Viterbi algorithm to obtain the most probable state sequence for FARC.
Figures~\ref{fig02}(a)--(c) illustrate the state classification for a
$\delta= 15$ day time window with $\{ X_n \}$, $\{ Y_n \}$ and
$\{ (X_n, \hsppp Y_n) \}$ as observations. As can be seen from
Figure~\ref{fig02}, the state classification approach detects even
\emph{small} and \emph{nonpersistent} changes. Further, the Viterbi algorithm
declares $51$ and $46$ of the $219$ time windows as \emph{Active} with
$\{ X_n \}$ and $\{ Y_n \}$, respectively, whereas the joint sequence
$\{ (X_n, \hsppp Y_n) \}$ results in the same classification as $\{ Y_n
\}$.
Table~\ref{table01} summarizes the number of time windows classified as
\emph{Active} for different $\delta$ values with $\{ (X_n, \hsppp
Y_n) \}$
as observations. This study also suggests that $\delta= 10$ to $15$
with the HMM approach optimally trading off the twin objectives of
detecting minor
spurts in the activity profile of FARC (larger $\mathsf{ N}_\mathsf{ spurt}$)
as well as minimizing the number of \emph{Active} state
classifications that
require further attention (smaller fractional activity $\mathsf{ f}$).

We now consider a more general two-parameter %(see Supplementary Part A)
hurdle-based geometric model for $\{ M_i \}$ that potentially allows the
joint sequence $\{ (X_n, \hsppp Y_n ) \}$ to result in better inferencing
on the states than either $\{X_n \}$ or $\{ Y_n \}$. As before, the
Baum--Welch algorithm is used to learn the underlying parameters with
different values of $\delta$ and $\{ (X_n, \hsppp Y_n ) \}$ as training
data. Table~\ref{table01} summarizes these parameter estimates and, as
was the case earlier, it can be seen that the estimates remain
stable across $\delta$.\ The parameter estimates are then used with the
Viterbi algorithm to infer the most probable state sequence [see
Figure~\ref{fig02}(d) for state classification in the $\delta= 15$ setting].
While Figure~\ref{fig02}(d) and Table~\ref{table01} show that the
classification with the hurdle-based model agrees with the simpler geometric
model for many $\delta$ values, in the setting of interest ($\delta=
7$ to
$15$ days), the hurdle-based model is more conservative in declaring an
\emph{Active} state. The four time windows of disagreement between the
hurdle-based geometric and geometric models for $\delta= 15$ correspond
to the boundary of minor spurts (time windows $11$, $64$, $191$ and $204$),
where the hurdle-based model is more conservative in declaring an \emph{Active}
state, whereas the geometric model is \emph{trigger-happy}.
\
\subsection{Shining Path}
\label{sec5b}

The second case study is the activity profile of Sendero Luminoso (more
popularly known as \emph{Shining Path}), a terrorist group in Peru. With
a focus on the sixteen-year period between 1981 and 1996, the RDWTI
reports a total of 163 incidents with a yearly breakdown of 10 incidents
in 1981, 7 in 1982, 10 in 1983, 7 in 1984, 3 in 1985, 12 in 1986, 19 in 1987,
7 in 1988, 18 in 1989,\vadjust{\goodbreak} 10 in 1990, 31 in 1991, 10 in 1992, 14 in 1993,
2 in 1994, 2 in 1995 and 1 in 1996. The choice of Shining Path and the
1981 to 1996 time period are motivated in supplementary material, Part
B [\citet{vasanthsuppB}].

%f5 #&#
\begin{figure}

\includegraphics{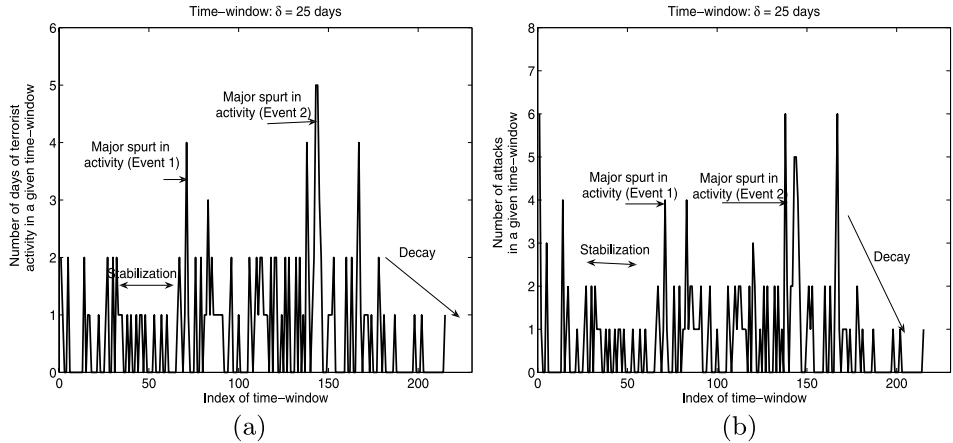}

\caption{\textup{(a)} Number of days of terrorist activity, and
\textup{(b)} Total number of attacks
in a $\delta= 25$ day time window as a function of time for Shining
Path.}\label{fig4}
\end{figure}

%f6 #&#
\begin{figure}

\includegraphics{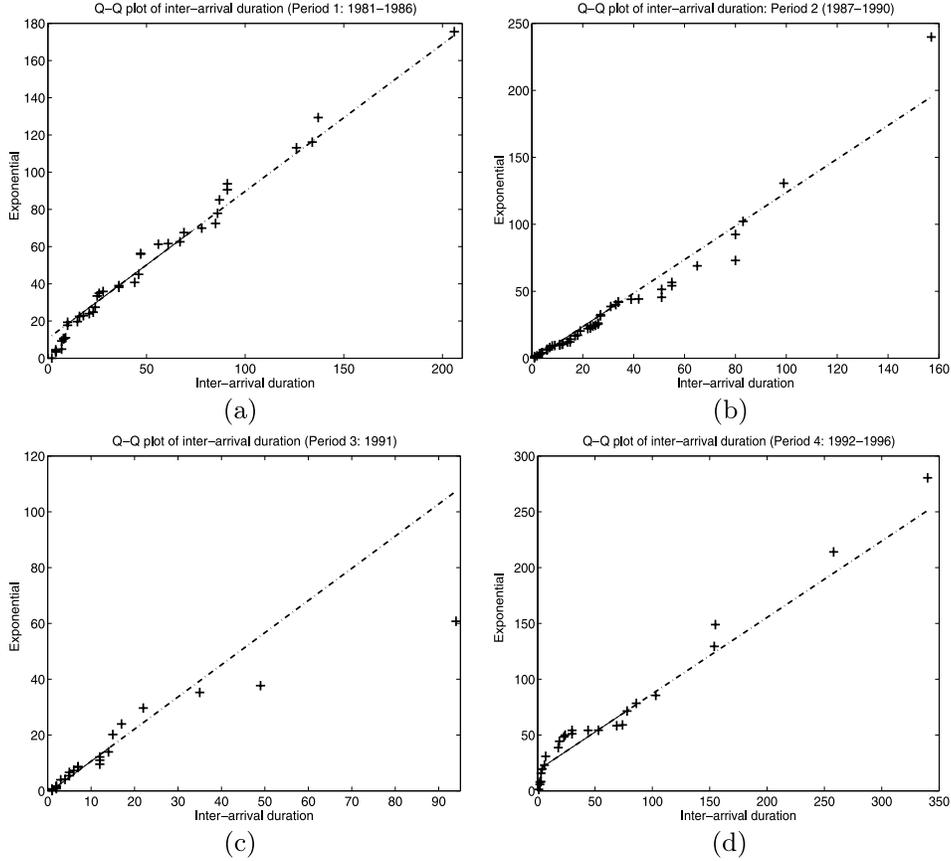}

\caption{Q--Q plots of %inter-arrival
duration between days of terrorist activity
with respect to a theoretical exponential distribution for the
period from: \textup{(a)} 1981 to 1986, \textup{(b)} 1987 to 1990,
\textup{(c)} 1991, and \textup{(d)} 1992
to 1996 corresponding to different stages of evolution of Shining
Path.}\label{fig5}
\end{figure}

In Figure~\ref{fig4}, we plot the number of days of terrorist activity and
the total number of attacks over a $\delta= 25$ day time window. In
Figure~\ref{fig5}, we test the fit of inter-arrival duration between
successive days of terrorist activity with respect to an %theoretical
exponential random variable with parameter $\widehat{\gamma}$
estimated as
in (\ref{est1}). As noted in supplementary material, Part B [\citet{vasanthsuppB}],
%Sec. \ref{secsuppb},
the evolution of Shining Path can be partitioned into four distinct phases.
%of a) rise and stabilization
%in the period from 1981 to 1986, b) resurgence and stabilization in the
%period from 1987 to 1990, c) heightened state of activity in 1991, and
%d) decay of the group in the period from 1992 to 1996 following the
%arrest
%of its leaders.
The inter-arrival duration in each phase can be distinctly
modeled as an exponential random variable and Figure~\ref{fig5} shows that
this partitioning is reasonable. As in the case with FARC, the exponential
random variable is a good first approximation, as the tails are not
well modeled with this random variable. Our numerical study leads to the
following estimates for $\gamma_{\bullet}$ in the four phases: (a)
$\widehat{\gamma} \approx0.0211$, (b) $\widehat{\gamma} \approx0.0344$,
(c)~$\widehat{\gamma} \approx0.0755$, (d) $\widehat{\gamma}
\approx0.0158$.
%which explains the geo-political undercurrents behind the evolution of
%Shining
%Path in these four phases.

\begin{figure}[b]\vspace*{6pt}

\includegraphics{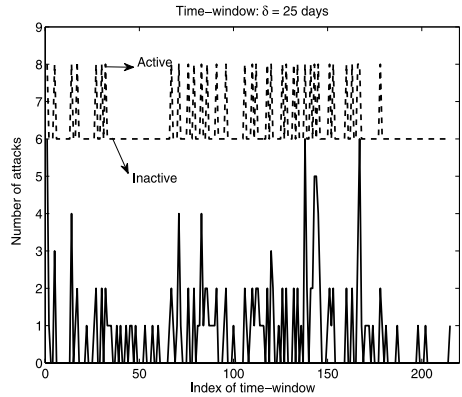}

\caption{State classification for Shining Path data ($\delta= 25$
days).}\label{fig6}
\end{figure}

As elucidated with the FARC data set, the data corresponding to Shining
Path is studied in the following experiment. Using the HMM approach with
the hurdle-based geometric model described in Section~\ref{sec4}, the
Baum--Welch algorithm results in parameter estimates as in Table~\ref{table03}
for different values of $\delta$. State classification via the Viterbi
algorithm using these estimates results in an \emph{Active}/\emph{Inactive}
classification for Shining Path, for example, Figure~\ref{fig6}
displays a typical
classification for $\delta= 25$. It is important to note that while four
distinct phases are identified in the evolution of Shining Path, only a
$d = 2$-state HMM is studied here. For the purpose of spurt/downfall
detection, even this coarse model is sufficient. As can be seen with FARC
data, even small and nonpersistent changes are detected by the HMM approach,
further confirming its usefulness.

%t2 #&#
\begin{table}
\caption{State classification of Shining Path with the hurdle-based
geometric model and $\{ (X_n, \hsppp Y_n) \}$ as observations for
different time window periods ($\delta$)}
\label{table03}
\tabcolsep=3pt
\begin{tabular*}{\textwidth}{@{\extracolsep{4in minus 4in}}lccccccc@{}}
\hline
& \multicolumn{4}{c}{\textbf{Parameters learned}} &&&\\[-6pt]
& \multicolumn{4}{c}{\hrulefill} &&&\\[-8pt]
\multicolumn{1}{c}{\multirow{2}{25pt}{$\bolds{\delta}$ \textbf{(in days)}}} &
\multicolumn{1}{c}{\multirow{1}{25pt}[-10pt]{\centering{$\bolds{\widehat{\gamma}_0}$}}} &
\multicolumn{1}{c}{\multirow{1}{25pt}[-10pt]{\centering{$\bolds{\widehat{\mu}_0}$}}} &
\multicolumn{1}{c}{\multirow{1}{25pt}[-10pt]{\centering{$\bolds{\widehat{\gamma}_1}$}}}&
\multicolumn{1}{c}{\multirow{1}{25pt}[-10pt]{\centering{$\bolds{\widehat{\mu}_1}$}}}&
\multicolumn{1}{c}{\multirow{3}{75pt}[10pt]{\centering{\textbf{Length of training set (}$\bolds{N}$ \textbf{time
windows)}}}}
&
\multicolumn{1}{c}{\multirow{3}{60pt}[10pt]{\centering\textbf{No. of \emph{Active}  states (}$\bolds{\mathsf{ N}_\mathsf{ spurt}}$ \textbf{time windows)}}}&
\multicolumn{1}{c@{}}{\multirow{3}{50pt}[10pt]{\centering\textbf{Fractional activity}
$\bolds{(\mathsf{ f} = \frac{ \mathsf{ N}_\mathsf{ spurt} \cdot\delta}{ {\cal
N}})}$}}\\\\
\hline
$20$ & $0.0262$ & $0.1026$ & $0.1000$ & $0.6667$ &
$268$ & $69$ & $0.2575$ \\
$25$ & $0.0264$ & $0.1019$ & $0.0800$ & $0.6667$ &
$215$ & $35$ & $0.1628$ \\
$28$ & $0.0264$ & $0.1019$ & $0.0714$ & $0.6667$ &
$192$ & $15$ & $0.0781$ \\
$30$ & $0.0262$ & $0.1026$ & $0.0667$ & $0.6667$ &
$179$ & $14$ & $0.0782$ \\
$32$ & $0.0202$ & $0.0703$ & $0.1072$ & $0.2255$ &
$168$ & $38$ & $0.2262$ \\
%$35$ & $0.0263$ & $0.1026$ & $0.0571$ & $0.6667$ &
%$153$ & $58$ & $0.3791$
\hline
\end{tabular*}
\end{table}
%

%f7 #&#

%
%State classification with Shining Path data ($\delta= 25$ days).}
%
%}

%s6 #&#
\section{Revisiting some of the modeling issues}
\label{sec6}
We now revisit some of the modeling issues that shed further light
on development of good models for the activity profile of terrorist
groups. %In this section, the primary focus is on the FARC dataset.

%s6.1 #&#
\subsection{Clustering of attacks}
\label{sec6a}
A central premise in terrorism modeling is that the
current activity of a terrorist group is influenced by its past
activity. One consequence of this premise is that the attacks perpetrated
by the group are \emph{clustered} [\citet{midlarsky,midlarsky1}]. Ripley's
$K(\cdot)$ function is a statistical tool for measuring the degree of
clustering (aggregatedness) or inhibition (regularity) in a point process
as a function of inter-point distance [\citet{diggle}]. Specifically, if
$\lambda$ is the intensity of the point process, $\lambda K(h)$ is the
expected number of other points within a distance $h$ of a randomly
chosen point of the process:
\begin{eqnarray*}
&&K(h) \triangleq\frac{1}{\lambda} \cdot \mathsf{ E} [\mbox{Number of other points within distance $h$}\\
&&\hspace*{125pt}\mbox{of a randomly chosen point}].
\end{eqnarray*}
Expressions for $K(h)$ can be derived for a number of stationary point
process models [\citet{dixon}]. For example, in the case of a
one-dimensional/tem\-poral
point process that is completely random (where points are distributed
uniformly and independently in time), it can be shown that $K(h) = 2h$.
A~two-dimensional complete random spatial point process leads to $K(h) =
\pi h^2$.
In the context of an activity profile, $K(h)$ can be estimated as
%
%
%e6.1 #&#
\begin{equation}\label{eqripley1}
\widehat{K}(h) %& = &
= \frac{1}{ \widehat{\lambda} \cdot N} \sum
_{i} \sum_{ j \neq i} \indic \bigl(
|t_i - t_j| \leq h \bigr) %\nonumber
= \frac{1} { ( \widehat{\lambda})^2 \cdot{\cal N}} \sum_{i} \sum
_{ j \neq i} \indic \bigl( |t_i - t_j|
\leq h \bigr),\hspace*{-35pt} %\nonumber
\end{equation}
where $t_i$ is the $i$th day of activity, $N$ is the number of
attacks in the observation period of ${\cal N}$ days (i.e.,
$[ t_1, \ldots, t_N ] = [ \mathrm{Day} \hspp1, \ldots,
\mathrm{Day} \hspp{\cal N}]$), and $\widehat{\lambda} = \frac{N} {
{\cal N} }$ is an estimate of the intensity (rate) of the
process. Significant deviations of $\widehat{K}(h)$ from $2h$ indicate
that the hypothesis of complete randomness becomes untenable with
observed data and more confidence is reposed on clustering [if
$\widehat{K}(h) \gg2h$] or inhibition [if $\widehat{K}(h) \ll2h$].

While the definition in (\ref{eqripley1}) assumes a homogenous point
process, extensions to an inhomogenous point process have also been
proposed [\citet{baddeley,veenschoenberg}], %marcon,ji_meng where the
where the point process is re-weighted by the reciprocal of the nonconstant
intensity function to offset the inhomogeneity. Further, to compensate
for edge effects due to points outside the observation period being
left out in the numerator of (\ref{eqripley1}), various edge-correction
estimators have also been proposed in the literature. Combining these
two facets, we have the following estimator\footnote{Note
that \citeauthor{porterwhite2012} propose a one-sided estimator for
$K(h)$ and they compare $\widehat{K}(h)$ with~$h$ (instead of $2h$)
to test for clustering/inhibition.} for $K(h)$:
%
%
%e6.2 #&#
\begin{equation}
\widehat{K}(h) = \frac{1}{{\cal N}} \sum_{i} \sum
_{j \neq i} \frac{
\indic ( |t_i - t_j| \leq h  ) }{ \widehat{p}_i \widehat{p}_j
w_{ij} }, %\nonumber
\label{eqripley2}
\end{equation}
where $\widehat{p}_i$ is the estimated probability of at least one attack
on $t_i$. In this work, we use
%modify the formula in (\ref{eqripley1}) by incorporating
an edge-correction factor $w_{ij}$ due to Ripley [\citet{cressie},
pages 616--618]
which reflects the proportion of the period centered at $t_i$ and covering
the $j$th day of activity that is included in the observation period:
\begin{eqnarray*}
w_{ij} = \cases{ %
 1, & $\quad\mathrm{if}\ \hspp
t_1 + R \leq t_i \leq%{\cal N}
t_N - R,$
\vspace*{2pt}\cr
\displaystyle\frac{ %{\cal N}
t_N - t_i + R }{ 2 R}, &\quad$ \mathrm{if}\ \hspp t_i > \max ( %{\cal N}
t_N - R, \hspp t_1 + R ),$
\vspace*{2pt}\cr
\displaystyle\frac{ R + t_i - t_1}{2R}, &\quad $\mathrm{if}\ \hspp t_i < \min ( %{\cal N}
t_N - R, \hspp t_1 + R ),$
\vspace*{2pt}\cr
\displaystyle\frac{ %{\cal N}
t_N - t_1} {2R}, & \quad$\mathrm{if}\ \hspp%{\cal N}
t_N - R \leq
t_i \leq t_1 + R,$}
\end{eqnarray*}
where $R = |t_i - t_j|$.

\begin{figure}

\includegraphics{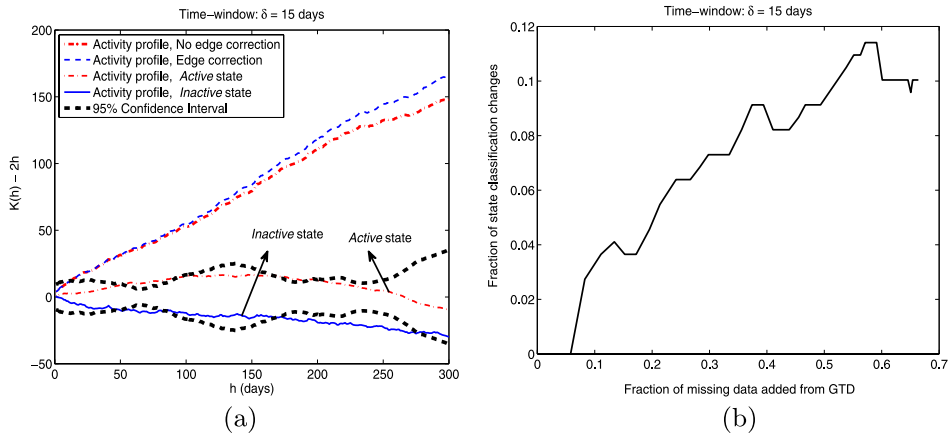}

\caption{\textup{(a)} $K(h)$ as a function of inter-point distance
$h$ for the activity
profile of FARC and under the two states of the HMM, \textup{(b)} fractional
change in state classification relative to addition of missing data
from GTD.} \label{figripley}
\end{figure}

In Figure~\ref{figripley}(a), we plot $K(h) - 2h$ computed as
in (\ref{eqripley2}) for the FARC data set (with and without
edge-correction) as a function of the inter-point distance $h$.
In line with the observation by \citet{porterwhite2012} for the
Indonesia/Timor-Leste data set, the plot here indicates that the
FARC data is also clustered since $K(h) - 2h \gg0$, thus motivating
the SEHM.

%f8 #&#

The main theme of this work, however, is that clustering is essentially
a reflection of state transitions and the activity sub-profile (sub-series)
of the group conditioned on a given state value is not clustered. To test
this hypothesis, we study the behavior of $K(h)$ for the sub-series from
the FARC data set corresponding to the \emph{Active} and \emph{Inactive}
states as classified by the methodology of Section~\ref{sec5a}. Since the
latent states of the group transition back and forth, an \emph{Active}
sub-series is constructed by patching together the group's activity
profile in the \emph{Active} state with the \emph{jump} random variable
between any two disjoint pieces of the activity profile modeled as
Poisson with parameter $\widehat{\lambda} = \frac{N} { {\cal N}}$. For
this sub-series, $K(h)$ is estimated using a formula analogous
to (\ref{eqripley2}), where $\widehat{p}_i$ and $w_{ij}$ are re-estimated
for the sub-series, and ${\cal N}$ is replaced with ${\cal N}_\mathsf{ act}$:
\[
 \hsp{\cal N}_\mathsf{act} = \min (\mbox{Number of days in \emph{Active} state, Length of \emph{Active} sub-series}).
\]
A confidence interval can also be constructed using the bootstrap
technique (resampling from the underlying data distribution). A similar
estimation process yields $K(h)$ and the corresponding confidence interval
for the \emph{Inactive} sub-series.

In contrast to the trends for the unclassified activity profile, both the
sub-series indicate an inhibitory behavior (mild inhibition
for the \emph{Active} state and stronger inhibition for the \emph{Inactive}
state) in Figure~\ref{figripley}(a). Further, $K(h) - 2h$ lies within the
$95 \%$ confidence interval (computed using a $1000$-point resampling)
for almost all $h$ values with the \emph{Active}
sub-series and for a significant fraction of $h$ values with the \emph
{Inactive}
sub-series. The stronger inhibition in the \emph{Inactive} state
should not be
entirely surprising since very few attacks happen in this state. The absence
of clustering in the activity profile conditioned on the latent state suggests
that the clustering of attacks can \emph{also} be explained as arising
due to
different intensity profiles in the different states. Thus, the HMM framework
offers a \emph{competing} approach to explain the clustering of attacks.

% prompts that the \emph{classical}
%assumption of clustering of attacks needs to be revisited and the
%carefully
%across many groups/datasets.

%t3 #&#
\begin{table}
\tabcolsep=1pt
\caption{Histogram of observed number of attacks per day for
FARC data with different model fits, $\delta=15$~days}
\label{table2}
\begin{tabular*}{\textwidth}{@{\extracolsep{\fill}}lccccccc@{}}
\hline
 &  &  & &  &  & \textbf{Hurdle-based}&\textbf{Hurdle-based}
\\
\textbf{No. attacks} & \textbf{Obs.} & \textbf{Poisson} & \textbf{Shifted zeta}&
\textbf{Geomet.} & \textbf{P\`{o}lya} & textbf{zeta}&
\textbf{geomet.}\\

 \hline
\multicolumn{8}{@{}l}{(\emph{Inactive} state)}\\
 $0$ & $2420$ & $2421$ & $2470$ & $2430$ & $2421$ & $2420$ & $2421$ \\ %
$1$ & $227$ & $225$ & $144$ & $207$ & $225$ & $229$ & $226$ \\ %\hline
$2$ & $9$ & $11$ & $27$ & $18$ & $11$ & $7$ & $10$ \\ %\hline
$3$ & $1$ & $0$ & $8$ & $2$ & $0$ & $1$ & $0$ \\ %\hline
$4$ & $0$ & $0$ & $4$ & $0$ & $0$ & $0$ & $0$ \\ %\hline
$> 4$ &
$0$ & $0$ & $4$ & $0$ & $0$ & $0$ & $0$ \\ %\hline
AIC & & $\textbf{1690.34}$ & $1772.81$ & $1696.74$ & $1692.32$ &
$1692.58$ & $1691.86$ \\ %\hline
Parameter & & $0.0933$ & $4.105$ & $0.0854$ &
$\widehat{r}_0 = 24.4749$, &
$\widehat{\gamma}_0 = 0.0892$, &
$\widehat{\mu}_0 = 0.0444$, \\
Estimate & & & & &
$\widehat{y}_0 = 0.0038$ & $\widehat{y}_0 = 5.10$ &
$\widehat{\gamma}_0 = 0.0892$
\\[3pt]% \hline\hline
%No. attacks & Obs. & Poisson & Shifted &
%Geomet. & P\`{o}lya & Hurdle- & Hurdle- \\
% & & & Zeta & & & Based & Based \\
% & & & & & & Zeta & Geomet. \\ \hline
\multicolumn{8}{@{}l}{(\emph{Active} state)}\\
$0$ & $384$ & $359$ & $455$ & $404$ & $389$ & $384$ & $384$ \\ %\hline
$1$ & $174$ & $202$ & $87$ & $144$ & $160$ & $189$ & $171$ \\ %\hline
$2$ & $46$ & $57$ & $33$ & $52$ & $56$ & $31$ & $52$ \\ %\hline
$3$ & $19$ & $11$ & $16$ & $19$ & $17$ & $11$ & $16$ \\ %\hline
$4$ & $4$ & $1$ & $9$ & $7$ & $6$ & $5$ & $5$ \\ %\hline
$> 4$ &
$3$ & $0$ & $30$ & $4$ & $2$ & $10$ & $2$ \\ %\hline
AIC & & $1313.88$ & $1416.88$ & $1291.73$ & $1288.85$ & $1308.09$ &
$\textbf{1287.11}$ \\ %\hline
Parameter & & $0.5651$ & $2.40$ & $0.3611$
& $\widehat{r}_1 = 1.4834$, & $\widehat{\gamma}_1 = 0.3905$, &
$\widehat{\mu}_1 = 0.3090$,
\\
Estimate & & & & &
$\widehat{y}_1 = 0.2759$ &
$\widehat{y}_1 = 2.61$ &
$\widehat{\gamma}_1 = 0.3905$
\\
\hline
\end{tabular*}\vspace*{-6pt}
\end{table}

%s6.2 #&#
\subsection{Model for observation densities}
\label{sec6b}
We now develop simple models for the observation density under the
two states. For this, we study the goodness of fit captured by the
AIC for several models with support on the nonnegative integers to
describe data from FARC. In Table~\ref{table2}, we present the histogram
of the number of days with $\ell$ ($\ell= 0,1, \ldots $) attacks per
day for FARC data. Applying the state classification algorithm described
in Section~\ref{sec4} with $\delta= 15$, FARC stays in the \emph{Inactive}
state for $2657$ days and in the \emph{Active} state for $630$ days. Also,
presented in the same table are the (rounded-off) expected number of days
with $\ell$ attacks for the six models described in supplementary material, Part A,
along with the AIC for these model fits. %Table \ref{table3} presents
%the
%corresponding histograms
While the corresponding data for $\delta= 10$ is not presented here due
to space constraints, FARC stays in the \emph{Inactive} state for $2577$
days and in the \emph{Active} state for $710$ days in this setting.

The ML parameter estimates for all the six models remain
robust as $\delta$ is varied, which is in conformity with the stability
of the converged Baum--Welch parameter estimates with $\delta$ (see
Tables~\ref{table01} and \ref{table03}). Further, from
Table~\ref{table2}, %and \ref{table3},
in the \emph{Inactive} state, it is
seen that all the models except the shifted zeta result in comparable
fits. Specifically, the hurdle-based geometric model differs from the
observed histogram in only one day and results in the second lowest AIC
value. On the other hand, in the \emph{Active} state, the hurdle-based
geometric model produces the best fit with only the P\`{o}lya model
resulting in a comparable fit. In this setting, the one-parameter models
overestimate either the tail or the days of no activity, while the
hurdle-based zeta produces a heavier tail than what the data exhibits.
In fact, the poorest fit in either state is obtained with the shifted zeta
suggesting that a heavy tail may not always be necessary. In contrast, the
Indonesia/Timor-Leste data studied in \citet{porterwhite2012} exhibits
several extreme values (e.g., days with $36, 11$ and $10$ attacks) and the
authors observe that a self-exciting hurdle-based zeta model captures the
heavy tails much better than other models. The FARC data set used here shows
a maximum of $7$ attacks per day, whereas the Shining Path %
%the data from GTD on Shining Path, which is more richer, is not
%expected to
%conform to the trend from RDWTI.}
data set shows a maximum of $3$ attacks per day. Even simple nonself-exciting
models are enough to capture these data sets well.

This study suggests the following:
\begin{itemize}
\item
If parsimony of the model is of critical importance, the geometric
distribution serves as the best one-parameter model with the Poisson/shifted
zeta models either under-estimating or overestimating the number of
days with no
activity in the \emph{Active} state.

\item
If parsimony is not a critical issue and the data does not have (or has
very few) extreme values, the hurdle-based geometric model serves as the
best/near-best model in either state.

\item
However, if the data has several extreme values [as seen
in \citet{porterwhite2012}], the self-exciting hurdle model offers
the best model fit, albeit at the expense of learning several model
parameters.
\end{itemize}

%s6.3 #&#
\subsection{Inter-arrival duration}
\label{sec6c}
We now study the efficacy of the HMM framework by testing the goodness
of fit
of the theoretical exponential random variable with respect to the
inter-arrival duration between days of terrorist activity in either
state. To avoid estimating the rate parameters of the exponential
random variables from the data (which complicates the hypothesis tests),
we use the fact from \citet{seshadri} that if $y_1, \ldots, y_m$
are i.i.d. exponential random
variables (with a given rate parameter), then
%
%
%e6.3 #&#
\begin{equation}
z_j = \frac{ \sum_{ i = 1}^j y_i } { \sum_{i= 1}^m y_i}, \qquad j = 1, \ldots, m %\nonumber
\label{transfeq}
\end{equation}
are i.i.d. uniformly distributed in $[0, 1]$. We then use a
Kolmogorov--Smirnov (KS) test to study the fit between the empirical
cumulative distribution function (CDF) of $z_j$ and the uniform
CDF [\citet{durbin}].

The KS test-statistic (denoted as $\mathsf{ KS}_{\bullet}$) and the
critical value for the test (denoted as $K_{\alpha}$) corresponding
to a significance level $\alpha$ and computed using the standard
asymptotic formula are given as
\begin{eqnarray*}
\mathsf{ KS}_{\bullet} & = & \max_{x} \Biggl| \frac{1}{ n_{\bullet} }
\sum_{i = 1}^{ n _{\bullet} } \indic \bigl(
z_i^{\bullet} \leq x \bigr) - x \Biggr|,
\\
K_{\alpha} & = & \sqrt{ \frac{ - ({1}/{2}) \log (
{\alpha}/{2}  )} {n_{\bullet} } },
\end{eqnarray*}
where $n_{\bullet}$ is the number of samples and $\{ z_i^{\bullet} \}$
are the transformed samples computed using (\ref{transfeq}) in either
state. The results of applying the KS test to the two states are
presented in
Table~\ref{table4}. From this table, it is clear that the samples in
either state fit the theoretical exponential assumption very accurately
with the exponential model rejected in the \emph{Active} (and \emph{Inactive})
state(s) if the significance level $\alpha$ exceeds $\alpha> 0.3763$
(and $\alpha> 0.7905$), respectively.

%t4 #&#
\begin{table}
\caption{KS test-statistic in the two states for FARC data set}
\label{table4}
%{}
%
\begin{tabular*}{\textwidth}{@{\extracolsep{\fill}}lcc@{}}
\hline
%|c||c|c|} \hline
& \multicolumn{1}{c}{\textbf{\emph{Active}}} & \multicolumn{1}{c@{}}{\textbf{\emph{Inactive}}} \\
\hline
Total number of samples & $245$ & $236$
\\ %\hline
No. of samples with $\Delta T_k \leq\beta$ &
$n_\mathsf{ A} = 234, \hspp\beta= 15$ &
$n_\mathsf{ I} = 192, \hspp\beta= 20$ \\
KS statistic & $0.0597$ & $0.0492$ \\ %\hline
$p$-value & $0.3763$ & $0.7905$ \\
\hline
\end{tabular*}
\end{table}

In a recent analysis of FARC activities using the MIPT database %
%MIPT database is comparable with RDWTI, but currently unavailable.}
over the time period 1998 to 2005, \citet{clauset2012} hypothesize
that the inter-arrival duration between successive days of attacks
``decreases consistently, albeit stochastically'' with the cumulative
number of events FARC has carried out---a~measure of the group's
experience. Our initial studies indicate that while this hypothesis holds
empirically true for $k \leq25$ attacks that encompass the period
January~1
to June 27, 1998, it consistently increases through the subsequent period
lasting till March 10, 2000. Note that this is the precise time period of
increased U.S. funding to combat FARC and the drug economy through
\emph{Plan Colombia} [\citet{cipaid}] and a mean increase in the time
to the
next day of activity indicates an impact of counter-terrorism
efforts. %by Colombian military, para-military and police personnel.
The following period through August~13, 2004 indicates a reversed trend
of consistent decrease, suggesting that the\vadjust{\goodbreak} organizational dynamics of FARC
had ``adjusted'' to the new reality of combat with the establishment. As
seen earlier, such distinct changes in the organizational dynamics (associated
with spurts and downfalls) are quickly identified by the approaches proposed
in this work.

%s6.4 #&#
\subsection{Robustness of proposed approach to missing data}
\label{sec6d}
We now study the robustness of the proposed spurt detection approach in
terms of
state classification to attacks that are not available in the
database. Toward this goal, we treat the FARC data from RDWTI
over the 1998 to 2006 period as the baseline data set and add one
missing day of activity per year from the GTD in a \emph{sequential}
manner and revisit state classification with the enhanced data set.
Specifically, the fraction of missing data added in the $j$th (sequential)
step is the ratio of the difference between new and old attacks and the
baseline, and is defined as
\[
\mathrm{Frac.\ Missing\ Data}(j) \triangleq
\frac{ \sum_{ i = 1}^{ {\cal N}} \widetilde{M}_i^j - M_i } {
\sum_{ i = 1}^{ {\cal N}} M_i }, %\frac{ \mathrm{No.} \hspp\mathrm{of} \hspp\mathrm{new} \hspp
%{ \mathrm{No.} \hspp\mathrm{of} \hspp\mathrm{old}
\]
where $\widetilde{M}_i^j$ is the number of attacks on the $i$th day with
data addition. Applying the state estimation algorithm proposed in
Section~\ref{sec4} to this new data set, let $\widehat{s}_{i, \hsppp j}
^{ \mathsf{ new} } \in\{ 0, 1\}$ denote the estimated state value on the
$i$th day ($i = 1, \ldots, {\cal N}$). The fractional change in state
classification is then defined as\looseness=-1
\[
\mathrm{Frac.\ State\ Classification\ Changes}(j)
\triangleq \frac{ \sum_{ i = 1}^{ {\cal N}}  | \widehat{s}_{i, \hsppp j} ^{
\mathsf{ new} }
- \widehat{s}_{i }  | } { {\cal N} }, %\frac{ \mathrm{No.} \hspp\mathrm{of} \hspp\mathrm{new} \hspp
%{ \mathrm{No.} \hspp\mathrm{of} \hspp\mathrm{old}
\]\looseness=0
with $ \widehat{s}_{i }$ denoting the state classification on the
$i$th day
with the baseline data set.

In Figure~\ref{figripley}(b), $\mathrm{Frac.}\ \mathrm{Missing}\ \mathrm{Data}(j)$ is plotted as a function of
$\mathrm{Frac.}\break \mathrm{State} \mathrm{Classification}\ \mathrm{Changes}(j)$. In general, combining
terrorism information from two different databases with a clear dichotomy
in terms of data collection standards (criteria for inclusion and noninclusion
of events, source material used, etc.) can introduce a systematic bias in
terrorism trends. % \citep{lafree2}.
Despite this anomaly, it is clear from
Figure~\ref{figripley}(b) that the proposed approach is remarkably robust
to a small amount of missing data. For example, the addition
of an $\approx5\%$ more data to the baseline data set leads to
essentially no
changes in state classification with the baseline data set. On the other
extreme, big additions of even up to $65 \%$ more data result in only an
$\approx10 \%$ mismatch in state classifications.

%s6.5 #&#
\subsection{Comparing the TAR, SEHM and HMM frameworks}
\label{sec6e}
It is important to compare the proposed HMM framework with the existing
TAR and SEHM frameworks in terms of the models' explanatory and
predictive powers. While
this comparison requires a careful study of performance across data
sets (and
is the subject of ongoing work), we now provide initial results in this
direction. We study both\vadjust{\goodbreak} models' ability to explain the %inter-attack
%duration
times to the subsequent day of activity $\{ \Delta T_1^n \}$ and their ability
to predict %the time to the next day of activity,
$\Delta T_{n+1}$ given $\{ \Delta T_1^n \}$. We validate both models
with the
FARC data set and the Indonesia/Timor-Leste data set\footnote{The
Indonesia/Timor-Leste data set from GTD extracted in January 2013 consists
of $291$ attacks over $165$ unique event days for the $1/1/1994$ to $12/31/2000$
period. The corresponding data set in \citet{porterwhite2012} consists of
$250$ attacks over $158$ unique event days. The discrepancy can be explained
as the addition of attacks to the GTD since the work of \citet
{porterwhite2012}.
Nevertheless, this discrepancy is not serious since we learn the model
parameters
for the enhanced data set from scratch.} studied by \citet{porterwhite2012}.

%t5 #&#
\begin{table}
\tabcolsep=0pt
\caption{Comparison between AIC and SMAPE scores with SEHM and HMM for
FARC and Indonesia/Timor-Leste data sets}
\label{tablemodels}
\fontsize{7.6}{10}{\selectfont{
\begin{tabular*}{\textwidth}{@{\extracolsep{4in minus 4in}}lcclcclcclcc@{}}
\hline
\multicolumn{6}{c}{ $\bolds{\mathsf{AIC}}$ } &
\multicolumn{6}{c}{ $\bolds{\mathsf{ SMAPE}}$ }
\\[-6pt]
\multicolumn{6}{c}{\hrulefill} &
\multicolumn{6}{c@{}}{\hrulefill}
\\
\multicolumn{3}{c}{$\bolds{\mathsf{ FARC}}$ } &
\multicolumn{3}{c}{ $\bolds{\mathsf{ Indonesia/Timor\mbox{-}Leste}}$ } &
\multicolumn{3}{c}{ $\bolds{\mathsf{ FARC}}$ } &
\multicolumn{3}{c}{ $\bolds{\mathsf{ Indonesia/Timor\mbox{-}Leste}}$ }
\\[-6pt]
\multicolumn{3}{c}{\hrulefill} &
\multicolumn{3}{c}{\hrulefill} &
\multicolumn{3}{c}{\hrulefill} &
\multicolumn{3}{c}{\hrulefill}\\
$\bolds{n}$ & $\bolds{\mathsf{ SEHM}}$ & $\bolds{\mathsf{ HMM}}$ & $\bolds{n}$ & $\bolds{\mathsf{ SEHM}}$ & $\bolds{\mathsf{ HMM}}$ &
$\bolds{n}$ & $\bolds{\mathsf{ SEHM}}$ & $\bolds{\mathsf{ HMM}}$ & $\bolds{n}$ & $\bolds{\mathsf{ SEHM}}$ & $\bolds{\mathsf{ HMM}}$
\\
\hline
$100$ & \phantom{0}$671.68$ & \phantom{0}$671.06$ & $100$ & \phantom{0}$723.78$ & \phantom{0}$729.47$ &
$100$ & $46.27 \%$ & $52.78 \%$ & $100$ & $46.33 \%$ & $43.32 \%$
\\
$200$ & $1117.40$ & $1112.07$ & $165$ & $1091.78$ & $1116.92$ &
$150$ & $42.95 \%$ & $35.75 \%$ & $125$ & $45.47 \%$ & $41.89 \%$
\\
$300$ & $1521.93$ & $1521.36$ & $200$ & $1283.08$ & $1305.27$ &
$200$ & $40.40 \%$ & $35.61 \%$ & $150$ & $42.84 \%$ & $38.75 \%$
\\
$400$ & $2127.55$ & $2121.81$ & $250$ & $1589.43$ & $1615.87$ &
$250$ & $40.09 \%$ & $38.14 \%$ & $175$ & $45.23 \%$ & $38.00 \%$
\\
$450$ & $2333.88$ & $2327.02$ & $300$ & $2018.92$ & $2041.35$ &
$300$ & $39.92 \%$ & $37.35 \%$ & $200$ & $43.46 \%$ & $33.99 \%$
\\
\hline
\end{tabular*}}}
\end{table}

The different baseline and self-exciting models considered
in \citet{porterwhite2012} are used to model $\{ \Delta T_1^n \}$ for
both data sets. The ${\tt fmincon}$ function in MATLAB is used to
learn model parameters that maximize the likelihood function
[see \citet{porterwhite2012}, equation (8)]. It turns out that a
four-parameter model (one parameter for the trend component and three
parameters for the negative binomial self-exciting component) is a good
model for both data sets. Table~\ref{tablemodels} shows the AIC
comparison between this four-parameter
SEHM and the four-parameter HMM for the two data sets. The results
suggest that from an explanatory
viewpoint, the SEHM is a better model than the HMM for the
Indonesia/Timor-Leste data set, whereas the HMM is better than the SEHM
for the FARC data set.

For comparing the predictive powers, model parameters are learned with
$\{ \Delta T_1^n \}$ as training data, and a conditional mean estimator
of the form $\widetilde{\Delta} T_{n+1} = \mathsf{ E}  [ \Delta
T_{n+1} |
\Delta T_1^n  ]$ is used for prediction. For the HMM framework,
it can
be checked that
\begin{eqnarray*}
\widetilde{\Delta} T_{n+1} |_\mathsf{ HMM} %& = &
&=& \sum
_{i = 0}^1 \mathsf{ P} \bigl(Q_{n+1} = i| \Delta
T_1^n \bigr) \mathsf{ E} [\Delta T_{n+1} |
Q_{n+1} = i ]\\ %\nonumber\\ & \stackrel{(a)}{=} &
&\stackrel{{\mathrm{(a)}}} {=}& \sum
_{i = 0}^1 \beta_i \mathsf{ E} [\Delta
T_{n+1} | Q_{n+1} = i ],
\end{eqnarray*}
where (a) follows from straightforward computations with
\[
\beta_i = \frac{ \sum_j \alpha_n(j) \mathsf{ P}(Q_{n+1} = i|Q_n = j)} {
\sum_j \alpha_n(j) }
\]
and $\alpha_n(j) = \mathsf{ P}(\Delta T_1^n,
Q_n = j)$
is updated via the forward procedure [\citet{rabiner}]. For the SEHM
framework, from (\ref{sehmeqn}), we have
\[
\widetilde{\Delta} T_{n+1} |_\mathsf{ SEHM} = %& = &
\frac{1} {1 - e^{- (B_n + \operatorname{SE}_n({\cal H}_{n-1}) )} }.
\]
For the sake of comparison, %purposes,
we also use a sample mean estimator as a baseline:
\[
\widetilde{\Delta} T_{n+1} |_\mathsf{ Baseline}  =  \frac{1}{n}
\sum_{i = 1}^n \Delta T_i.
\]

\begin{figure}

\includegraphics{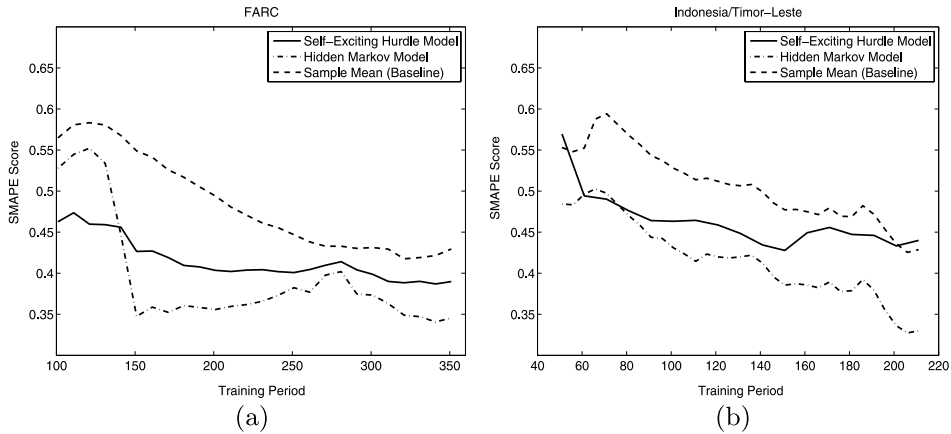}

\caption{SMAPE scores for the three models with \textup{(a)} FARC data and \textup{(b)}
Indonesia/Timor-Leste
data.}\label{fig10}
\end{figure}

To compare the three prediction algorithms, we use the Symmetric Mean Absolute
Percentage Error (SMAPE) score, defined as
%
%e6.4 #&#
\begin{equation}
\mathsf{ SMAPE} \triangleq\frac{1}{N} \sum_{i = 1}^N
\biggl\llvert \frac{
\Delta T_i -
\widetilde{\Delta} T_i }{ \Delta T_i + \widetilde{\Delta} T_i } \biggr\rrvert.
\end{equation}
Recall that the SMAPE score captures the relative error in prediction and
is a number between $0 \%$ and $100 \%$ with a smaller value indicating
a better
prediction algorithm. The SMAPE scores of the time to the next day of activity
for the three estimators (HMM, SEHM and baseline) are plotted as a function
of the training period for model learning in Figure~\ref{fig10}(a) for
the FARC data set
and in Figure~\ref{fig10}(b) for the Indonesia/Timor-Leste data set.
It can be
seen from Figure~\ref{fig10} that for both the data sets, the HMM
results in a
better prediction than the SEHM and the baseline provided the training period
is long to ensure accurate model learning for the HMM.

%f9 #&#

%%\begin{center}
%%\includegraphics[height=2.6in,width=3in] {fig/rel_aic_gain1.eps}
%{smape_scores_farc_fig4_bw.eps}
%&
%{smape_scores_timor_leste4_bw.eps}
%{} (a) & {} (b)
%%\end{minipage}
%
%

We now provide a qualitative comparison between the TAR, SEHM and HMM
frameworks.
While all the three models assume that the current observation/activity
is dependent
on the past history, the models differ in how this dependence is
realized. In
particular, in the TAR model, the current observation is explicitly
dependent on
the past observations along with (possibly) the impact from other independent
variables corresponding to certain geopolitical events/\emph
{interventions}. On
the other hand, in the SEHM, the probability of an attack is enhanced by
the history of the group according to the formula:
\[
\frac{ \mathsf{ P}(M_i > 0 | {\cal H}_{i-1} )  |_\mathsf{ SEHM} } {
\mathsf{ P}(M_i > 0 | {\cal H}_{i-1} )  |_\mathsf{ Non\mbox{-}SEHM}} = 1 + \frac{ e^{-B_i} }{ 1 - e^{-B_i} } \cdot \bigl( 1 - e^{-\operatorname{SE}_i( {\cal H}_{i-1}) }
\bigr) \geq1.
\]
The HMM combines both these facets by introducing a hidden state sequence.
The state sequence depends explicitly on its most immediate past (one-step
Markovian structure), whereas the probability of an attack is enhanced based
on the state realization.

The TAR model and the HMM are similar from the viewpoint of regime switching,
as these features are modeled explicitly. However, the mechanism of regime
switching is different in the two cases: the former assumes a change in the
auto-regressive process, whereas the latter assumes a state transition in
the HMM. The SEHM also incorporates a switch between states (induced by the
self-exciting component), but this switch is more of an implicit
feature of
the model rather than an explicit component.

More importantly, the TAR model considers global terrorism trends
rather than
trends constrained to a specific region or a specific group. Similarly, the
Indonesia/Timor-Leste data set considered by \citet{porterwhite2012}
is a
collation of \emph{all} attacks in Indonesia and Timor-Leste from
diverse groups
with significantly different \emph{Intentions} and \emph
{Capabilities} profiles
such as \emph{Dar-ul-Islam}, \emph{Gerakan Aceh Merdeka}, \emph
{Jemaah Islamiyah},
etc. On the other hand, the FARC data set considered here is
exclusively the
action of the many sub-groups of FARC. This distinction between
activity across
groups in a specific region versus group-based activity could explain
why the
HMM leads to a better model fit for the FARC data set relative to the
SEHM. This
logic also suggests that the HMM may be a poorer model for regional/global
trends. This hypothesis deserves a more careful study and is the
subject of
current work.

%s7 #&#
\section{Concluding remarks}
\label{sec7}
This work develops a HMM framework to model the activity profile of
terrorist groups. Key to this development is the hypothesis that the
current activity of the group can be captured completely by certain
states/attributes of the group, instead of the entire past history of
the group.
In the simplest example of the proposed framework, the group's activity
is captured by a $d = 2$ state HMM with the states reflecting a low state
of activity (\emph{Inactive}) and a high state of activity (\emph{Active}),
respectively. In either state, the days of activity are modeled as a
discrete-time Poisson point process with a hurdle-based geometric model
being a good fit for the number of attacks per day. While more general
models can be considered, even the simplest framework is sufficient for
detecting spurts and downfalls in the activity profile of many groups of
interest. Our results show that the HMM approach provides a competent
alternate modeling framework to the TAR and SEHM approaches, both in terms
of explanatory and predictive powers.

Fruitful directions to explore in the future include development of more
refined models for the activity profile (such as hierarchical HMMs) that
incorporate heavy tails and extreme outliers commonly observed in terrorism
data. A systematic comparison between the TAR model, SEHM and HMM and a
possible bridge between these classes will also be of interest. In
terms of
inferencing, nonlinear filtering approaches such as particle filters
are of
importance in practice. Given the intensive nature of data collection that
is common for studies of this nature, it would be of interest in developing
broad trends and trade-offs in quantitative terrorism studies with a large
set of groups from different ideological proclivities.


\begin{supplement}[id=suppA]
\sname{Supplement A}
\stitle{Information on models for the number of attacks
per day studied in this work}
\slink[doi]{10.1214/13-AOAS682SUPPA} %[doi,text={...}] - jei reikia suskaldyti doi
\sdatatype{.pdf}
\sfilename{aoas682\_suppa.pdf}
\sdescription{This section derives the ML and Baum--Welch estimate of model
parameter(s) under the geometric and hurdle-based geometric assumptions
on $\{ M_i \}$.}
\end{supplement}

\begin{supplement}[id=suppA]
\sname{Supplement B}
\stitle{Background information on FARC and shining path}
\slink[doi]{10.1214/13-AOAS682SUPPB} %[doi,text={...}] - jei reikia suskaldyti doi
\sdatatype{.pdf}
\sfilename{aoas682\_suppb.pdf}
\sdescription{This section motivates the choice of the terrorist
groups and
the corresponding time periods of interest that are the focus of this work.}
\end{supplement}

% imsref loaded by akundreckaite, 2013-10-18 13:38:35
% imsref loaded by akundreckaite, 2013-10-18 15:18:11
% imsref loaded by akundreckaite, 2013-10-21 15:03:05

%

\printaddresses


\begin{thebibliography}{38}
% BibTex style file: ims.bst, 2013-01-28
% Default style options (sort=0,type=number).
% Used options (sort=1,type=nameyear).

\bibitem[\protect\citeauthoryear{Baddeley, M{\o}ller and
  Waagepetersen}{2000}]{baddeley}
\begin{barticle}[mr]
\bauthor{\bsnm{Baddeley},~\bfnm{A.~J.}\binits{A.~J.}},
  \bauthor{\bsnm{M{\o}ller},~\bfnm{J.}\binits{J.}} \AND
  \bauthor{\bsnm{Waagepetersen},~\bfnm{R.}\binits{R.}}
(\byear{2000}).
\btitle{Non- and semi-parametric estimation of interaction in inhomogeneous
  point patterns}.
\bjournal{Stat. Neerl.}
\bvolume{54}
\bpages{329--350}.
\bid{doi={10.1111/1467-9574.00144}, issn={0039-0402}, mr={1804002}}
\bptok{imsref}%
\end{barticle}
\endbibitem

\bibitem[\protect\citeauthoryear{Cho et~al.}{2013}]{cho2013}
\begin{bmisc}[author]
\bauthor{\bsnm{Cho},~\bfnm{Y.~S.}\binits{Y.~S.}},
  \bauthor{\bsnm{Galstyan},~\bfnm{A.}\binits{A.}},
  \bauthor{\bsnm{Brantingham},~\bfnm{P.~J.}\binits{P.~J.}} \AND
  \bauthor{\bsnm{Tita},~\bfnm{G.}\binits{G.}}
(\byear{2013}).
\bhowpublished{Latent point process models for spatial-temporal networks.
  Available at \arxivurl{arXiv:1302.2671}}.
\bptok{imsref}%
\end{bmisc}
\endbibitem

\bibitem[\protect\citeauthoryear{Clauset and Gleditsch}{2012}]{clauset2012}
\begin{barticle}[pbm]
\bauthor{\bsnm{Clauset},~\bfnm{Aaron}\binits{A.}} \AND
  \bauthor{\bsnm{Gleditsch},~\bfnm{Kristian~Skrede}\binits{K.~S.}}
(\byear{2012}).
\btitle{The developmental dynamics of terrorist organizations}.
\bjournal{PLoS ONE}
\bvolume{7}
\bpages{e48633}.
\bid{doi={10.1371/journal.pone.0048633}, issn={1932-6203},
  pii={PONE-D-12-25296}, pmcid={3504060}, pmid={23185267}}
\bptok{imsref}%
\end{barticle}
\endbibitem

\bibitem[\protect\citeauthoryear{Clauset, Young and Gleditsch}{2007}]{clauset}
\begin{barticle}[author]
\bauthor{\bsnm{Clauset},~\bfnm{A.}\binits{A.}},
  \bauthor{\bsnm{Young},~\bfnm{M.}\binits{M.}} \AND
  \bauthor{\bsnm{Gleditsch},~\bfnm{K.~S.}\binits{K.~S.}}
(\byear{2007}).
\btitle{On the frequency of severe terrorist events}.
\bjournal{Journal of Conflict Resolution}
\bvolume{51}
\bpages{58--87}.
\bptok{imsref}%
\end{barticle}
\endbibitem

\bibitem[\protect\citeauthoryear{Cox and Isham}{1980}]{coxisham}
\begin{bbook}[mr]
\bauthor{\bsnm{Cox},~\bfnm{David~Roxbee}\binits{D.~R.}} \AND
  \bauthor{\bsnm{Isham},~\bfnm{Valerie}\binits{V.}}
(\byear{1980}).
\btitle{Point Processes}.
\bpublisher{Chapman \& Hall}, \blocation{London}.
\bid{mr={0598033}}
\bptok{imsref}%
\end{bbook}
\endbibitem

\bibitem[\protect\citeauthoryear{Cragin and Daly}{2004}]{rand}
\begin{bbook}[author]
\bauthor{\bsnm{Cragin},~\bfnm{K.}\binits{K.}} \AND
  \bauthor{\bsnm{Daly},~\bfnm{S.~A.}\binits{S.~A.}}
(\byear{2004}).
\btitle{{The Dynamic Terrorist Threat: An Assessment of Group Motivations and
  Capabilities in a Changing World}}.
\bpublisher{RAND Corporation}, \blocation{Santa Monica, CA}.
\bptok{imsref}%
\end{bbook}
\endbibitem

\bibitem[\protect\citeauthoryear{Cressie}{1991}]{cressie}
\begin{bbook}[mr]
\bauthor{\bsnm{Cressie},~\bfnm{Noel A.~C.}\binits{N.~A.~C.}}
(\byear{1991}).
\btitle{Statistics for Spatial Data}.
\bpublisher{Wiley}, \blocation{New York}.
\bid{mr={1127423}}
\bptok{imsref}%
\end{bbook}
\endbibitem

\bibitem[\protect\citeauthoryear{Diggle}{2003}]{diggle}
\begin{bbook}[author]
\bauthor{\bsnm{Diggle},~\bfnm{P.~J.}\binits{P.~J.}}
(\byear{2003}).
\btitle{{Statistical Analysis of Spatial Point Patterns}},
\bedition{2nd} ed.
\bpublisher{Edward Arnold}, \blocation{London}.
\bptok{imsref}%
\end{bbook}
\endbibitem

\bibitem[\protect\citeauthoryear{Dixon}{2002}]{dixon}
\begin{bincollection}[author]
\bauthor{\bsnm{Dixon},~\bfnm{P.~M.}\binits{P.~M.}}
(\byear{2002}).
\btitle{{Ripley's $K$ function}}.
In \bbooktitle{Encyclopedia of Environmetrics}
(\beditor{\bfnm{A.~H.}\binits{A.~H.}~\bsnm{El-Shaarawi}} \AND
  \beditor{\bfnm{W.~W.}\binits{W.~W.}~\bsnm{Piegorsc}}, eds.)
\bvolume{2}
\bpages{1796--1803}.
\bpublisher{Wiley}, \blocation{Chichester}.
\bptok{imsref}%
\end{bincollection}
\endbibitem

\bibitem[\protect\citeauthoryear{Dugan, LaFree and
  Piquero}{2005}]{duganlafreepiquero}
\begin{barticle}[author]
\bauthor{\bsnm{Dugan},~\bfnm{L.}\binits{L.}},
  \bauthor{\bsnm{LaFree},~\bfnm{G.}\binits{G.}} \AND
  \bauthor{\bsnm{Piquero},~\bfnm{A.}\binits{A.}}
(\byear{2005}).
\btitle{{Testing a rational choice model of airline hijackings}}.
\bjournal{Criminology}
\bvolume{43}
\bpages{1031--1066}.
\bptok{imsref}%
\end{barticle}
\endbibitem

\bibitem[\protect\citeauthoryear{Durbin}{1973}]{durbin}
\begin{bbook}[mr]
\bauthor{\bsnm{Durbin},~\bfnm{J.}\binits{J.}}
(\byear{1973}).
\btitle{Distribution Theory for Tests Based on the Sample Distribution
  Function}.
\bpublisher{SIAM},
  \blocation{Philadelphia, PA.}
\bid{mr={0305507}}
\bptok{imsref}%
\end{bbook}
\endbibitem

\bibitem[\protect\citeauthoryear{Enders and Sandler}{1993}]{enderssandler1993}
\begin{barticle}[author]
\bauthor{\bsnm{Enders},~\bfnm{W.}\binits{W.}} \AND
  \bauthor{\bsnm{Sandler},~\bfnm{T.}\binits{T.}}
(\byear{1993}).
\btitle{{The effectiveness of antiterrorism policies: A vector
  autoregression-intervention analysis}}.
\bjournal{The American Political Science Review}
\bvolume{87}
\bpages{829--844}.
\bptok{imsref}%
\end{barticle}
\endbibitem

\bibitem[\protect\citeauthoryear{Enders and Sandler}{2000}]{enderssandler2000}
\begin{barticle}[author]
\bauthor{\bsnm{Enders},~\bfnm{W.}\binits{W.}} \AND
  \bauthor{\bsnm{Sandler},~\bfnm{T.}\binits{T.}}
(\byear{2000}).
\btitle{{Is transnational terrorism becoming more threatening? A~time-series
  investigation}}.
\bjournal{Journal of Conflict Resolution}
\bvolume{44}
\bpages{307--332}.
\bptok{imsref}%
\end{barticle}
\endbibitem

\bibitem[\protect\citeauthoryear{Enders and Sandler}{2002}]{enderssandler2002}
\begin{barticle}[author]
\bauthor{\bsnm{Enders},~\bfnm{W.}\binits{W.}} \AND
  \bauthor{\bsnm{Sandler},~\bfnm{T.}\binits{T.}}
(\byear{2002}).
\btitle{{Patterns of transnational terrorism, 1970--1999: Alternative
  time-series estimates}}.
\bjournal{International Studies Quarterly}
\bvolume{2}
\bpages{145--165}.
\bptok{imsref}%
\end{barticle}
\endbibitem

\bibitem[\protect\citeauthoryear{Haugaard, Isacson and Olson}{2005}]{cipaid}
\begin{bmisc}[author]
\bauthor{\bsnm{Haugaard},~\bfnm{L.}\binits{L.}},
  \bauthor{\bsnm{Isacson},~\bfnm{A.}\binits{A.}} \AND
  \bauthor{\bsnm{Olson},~\bfnm{J.}\binits{J.}}
(\byear{2005}).
\bhowpublished{Erasing the lines: Trends in U.S. military programs with Latin
  America. Technical report, Center for International Policy, Washington, DC}.
\bptok{imsref}%
\end{bmisc}
\endbibitem

\bibitem[\protect\citeauthoryear{Hawkes}{1971}]{hawkes1971}
\begin{barticle}[mr]
\bauthor{\bsnm{Hawkes},~\bfnm{Alan~G.}\binits{A.~G.}}
(\byear{1971}).
\btitle{Spectra of some self-exciting and mutually exciting point processes.}
\bjournal{Biometrika}
\bvolume{58}
\bpages{83--90}.
\bid{issn={0006-3444}, mr={0278410}}
\bptok{imsref}%
\end{barticle}
\endbibitem

\bibitem[\protect\citeauthoryear{ITERATE}{2004}]{iterate}
\begin{bmisc}[author]
\borganization{ITERATE}
(\byear{2004}).
\bhowpublished{International terrorism: Attributes of terrorist events.
  Available at \url{http://www.icpsr.umich.edu/icpsrweb/ICPSR/studies/07947}.}
\bptok{imsref}%
\end{bmisc}
\endbibitem

\bibitem[\protect\citeauthoryear{LaFree and Dugan}{2007}]{lafree1}
\begin{barticle}[author]
\bauthor{\bsnm{LaFree},~\bfnm{G.}\binits{G.}} \AND
  \bauthor{\bsnm{Dugan},~\bfnm{L.}\binits{L.}}
(\byear{2007}).
\btitle{{Introducing the global terrorism database}}.
\bjournal{Terrorism and Political Violence}
\bvolume{19}
\bpages{181--204}.
\bptok{imsref}%
\end{barticle}
\endbibitem

\bibitem[\protect\citeauthoryear{LaFree, Morris and
  Dugan}{2010}]{lafreemorrisdugan}
\begin{barticle}[author]
\bauthor{\bsnm{LaFree},~\bfnm{G.}\binits{G.}},
  \bauthor{\bsnm{Morris},~\bfnm{N.~A.}\binits{N.~A.}} \AND
  \bauthor{\bsnm{Dugan},~\bfnm{L.}\binits{L.}}
(\byear{2010}).
\btitle{{Cross-national patterns of terrorism, comparing trajectories for
  total, attributed and fatal attacks, 1970--2006}}.
\bjournal{British Journal of Criminology}
\bvolume{50}
\bpages{622--649}.
\bptok{imsref}%
\end{barticle}
\endbibitem

\bibitem[\protect\citeauthoryear{Lewis et~al.}{2011}]{lewis}
\begin{barticle}[author]
\bauthor{\bsnm{Lewis},~\bfnm{E.}\binits{E.}},
  \bauthor{\bsnm{Mohler},~\bfnm{G.~O.}\binits{G.~O.}},
  \bauthor{\bsnm{Brantingham},~\bfnm{P.~J.}\binits{P.~J.}} \AND
  \bauthor{\bsnm{Bertozzi},~\bfnm{A.}\binits{A.}}
(\byear{2011}).
\btitle{{Self-exciting point process models of civilian deaths in Iraq}}.
\bjournal{Security Journal}
\bvolume{25}
\bpages{244--264}.
\bptok{imsref}%
\end{barticle}
\endbibitem

\bibitem[\protect\citeauthoryear{Lindberg}{2010}]{lindberg}
\begin{bmisc}[author]
\bauthor{\bsnm{Lindberg},~\bfnm{M.}\binits{M.}}
(\byear{2010}).
\bhowpublished{Factors contributing to the strength and resilience of terrorist
  groups. Grupo de Estudios Estrategicos (GEES) Publication.}
\bptok{imsref}%
\end{bmisc}
\endbibitem

\bibitem[\protect\citeauthoryear{Midlarsky}{1978}]{midlarsky}
\begin{barticle}[author]
\bauthor{\bsnm{Midlarsky},~\bfnm{M.~I.}\binits{M.~I.}}
(\byear{1978}).
\btitle{{Analyzing diffusion and contagion effects: The urban disorders of the
  1960s}}.
\bjournal{The American Political Science Review}
\bvolume{72}
\bpages{996--1008}.
\bptok{imsref}%
\end{barticle}
\endbibitem

\bibitem[\protect\citeauthoryear{Midlarsky, Crenshaw and
  Yoshida}{1980}]{midlarsky1}
\begin{barticle}[author]
\bauthor{\bsnm{Midlarsky},~\bfnm{M.~I.}\binits{M.~I.}},
  \bauthor{\bsnm{Crenshaw},~\bfnm{M.}\binits{M.}} \AND
  \bauthor{\bsnm{Yoshida},~\bfnm{F.}\binits{F.}}
(\byear{1980}).
\btitle{{Why violence spreads: The contagion of international terrorism}}.
\bjournal{International Studies Quarterly}
\bvolume{24}
\bpages{262--298}.
\bptok{imsref}%
\end{barticle}
\endbibitem

\bibitem[\protect\citeauthoryear{Mohler et~al.}{2011}]{mohler}
\begin{barticle}[mr]
\bauthor{\bsnm{Mohler},~\bfnm{G.~O.}\binits{G.~O.}},
  \bauthor{\bsnm{Short},~\bfnm{M.~B.}\binits{M.~B.}},
  \bauthor{\bsnm{Brantingham},~\bfnm{P.~J.}\binits{P.~J.}},
  \bauthor{\bsnm{Schoenberg},~\bfnm{F.~P.}\binits{F.~P.}} \AND
  \bauthor{\bsnm{Tita},~\bfnm{G.~E.}\binits{G.~E.}}
(\byear{2011}).
\btitle{Self-exciting point process modeling of crime}.
\bjournal{J. Amer. Statist. Assoc.}
\bvolume{106}
\bpages{100--108}.
\bid{doi={10.1198/jasa.2011.ap09546}, issn={0162-1459}, mr={2816705}}
\bptok{imsref}%
\end{barticle}
\endbibitem

\bibitem[\protect\citeauthoryear{Mueller and Stewart}{2011}]{muellerstewart}
\begin{bbook}[author]
\bauthor{\bsnm{Mueller},~\bfnm{J.}\binits{J.}} \AND
  \bauthor{\bsnm{Stewart},~\bfnm{M.~G.}\binits{M.~G.}}
(\byear{2011}).
\btitle{{Terrorism, Security, and Money: Balancing the Risks, Benefits, and
  Costs of Homeland Security}}.
\bpublisher{Oxford Univ. Press}, \blocation{London}.
\bptok{imsref}%
\end{bbook}
\endbibitem

\bibitem[\protect\citeauthoryear{Ogata}{1988}]{ogata1}
\begin{barticle}[author]
\bauthor{\bsnm{Ogata},~\bfnm{Y.}\binits{Y.}}
(\byear{1988}).
\btitle{{Statistical models for earthquake occurrences and residual analysis
  for point processes}}.
\bjournal{J. Amer. Statist. Assoc.}
\bvolume{83}
\bpages{9--27}.
\bptok{imsref}%
\end{barticle}
\endbibitem

\bibitem[\protect\citeauthoryear{Ogata}{1998}]{ogata}
\begin{barticle}[author]
\bauthor{\bsnm{Ogata},~\bfnm{Y.}\binits{Y.}}
(\byear{1998}).
\btitle{{Space--time point process models for earthquake occurrences}}.
\bjournal{Ann. Inst. Statist. Math.}
\bvolume{50}
\bpages{379--402}.
\bptok{imsref}%
\end{barticle}
\endbibitem

\bibitem[\protect\citeauthoryear{Porter and White}{2012}]{porterwhite2012}
\begin{barticle}[mr]
\bauthor{\bsnm{Porter},~\bfnm{Michael~D.}\binits{M.~D.}} \AND
  \bauthor{\bsnm{White},~\bfnm{Gentry}\binits{G.}}
(\byear{2012}).
\btitle{Self-exciting hurdle models for terrorist activity}.
\bjournal{Ann. Appl. Stat.}
\bvolume{6}
\bpages{106--124}.
\bid{doi={10.1214/11-AOAS513}, issn={1932-6157}, mr={2951531}}
\bptok{imsref}%
\end{barticle}
\endbibitem

\bibitem[\protect\citeauthoryear{Rabiner}{1989}]{rabiner}
\begin{barticle}[author]
\bauthor{\bsnm{Rabiner},~\bfnm{L.~R.}\binits{L.~R.}}
(\byear{1989}).
\btitle{{A tutorial on hidden Markov models and selected applications in speech
  recognition}}.
\bjournal{Proceedings of the IEEE}
\bvolume{77}
\bpages{257--286}.
\bptok{imsref}%
\end{barticle}
\endbibitem

\bibitem[\protect\citeauthoryear{Raghavan, Galstyan and
  Tartakovsky}{2012}]{vasantharxiv2012}
\begin{bmisc}[author]
\bauthor{\bsnm{Raghavan},~\bfnm{V.}\binits{V.}},
  \bauthor{\bsnm{Galstyan},~\bfnm{A.}\binits{A.}} \AND
  \bauthor{\bsnm{Tartakovsky},~\bfnm{A.~G.}\binits{A.~G.}}
(\byear{2012}).
\bhowpublished{Hidden Markov models for the activity profile of terrorist
  groups. Available at \arxivurl{arXiv:1207.1497v2}.}
\bptok{imsref}%
\end{bmisc}
\endbibitem

\bibitem[\protect\citeauthoryear{Raghavan, Galstyan and
  Tartakovsky}{2013a}]{vasanthsuppA}
\begin{bmisc}[author]
\bauthor{\bsnm{Raghavan},~\bfnm{V.}\binits{V.}},
  \bauthor{\bsnm{Galstyan},~\bfnm{A.}\binits{A.}} \AND
  \bauthor{\bsnm{Tartakovsky},~\bfnm{A.~G.}\binits{A.~G.}}
(\byear{2013}a).
\bhowpublished{Supplement to ``Hidden Markov models for the activity profile of
  terrorist groups.'' DOI:\doiurl{10.1214/13-AOAS682SUPPA}.}
\bptok{imsref}%
\end{bmisc}
\endbibitem

\bibitem[\protect\citeauthoryear{Raghavan, Galstyan and
  Tartakovsky}{2013b}]{vasanthsuppB}
\begin{bmisc}[author]
\bauthor{\bsnm{Raghavan},~\bfnm{V.}\binits{V.}},
  \bauthor{\bsnm{Galstyan},~\bfnm{A.}\binits{A.}} \AND
  \bauthor{\bsnm{Tartakovsky},~\bfnm{A.~G.}\binits{A.~G.}}
(\byear{2013}b).
\bhowpublished{Supplement to ``Hidden Markov models for the activity profile of
  terrorist groups.'' DOI:\doiurl{10.1214/13-AOAS682SUPPB}.}
\bptok{imsref}%
\end{bmisc}
\endbibitem

\bibitem[\protect\citeauthoryear{RDWTI}{}]{rdwti}
\begin{bmisc}[author]
\borganization{RDWTI}.
\bhowpublished{RAND database of worldwide terrorism incidents. Available at
  \url{http://www.rand.org/nsrd/projects/terrorism-incidents.html}.}
\bptok{imsref}%
\end{bmisc}
\endbibitem

\bibitem[\protect\citeauthoryear{Santos}{2011}]{santos}
\begin{barticle}[author]
\bauthor{\bsnm{Santos},~\bfnm{D.~N.}\binits{D.~N.}}
(\byear{2011}).
\btitle{{What constitutes terrorist network resiliency?}}
\bjournal{Small Wars Journal}
\bvolume{7}.
\bptok{imsref}%
\end{barticle}
\endbibitem

\bibitem[\protect\citeauthoryear{Seshadri, Csorgo and
  Stephens}{1969}]{seshadri}
\begin{barticle}[author]
\bauthor{\bsnm{Seshadri},~\bfnm{V.}\binits{V.}},
  \bauthor{\bsnm{Csorgo},~\bfnm{M.}\binits{M.}} \AND
  \bauthor{\bsnm{Stephens},~\bfnm{M.~A.}\binits{M.~A.}}
(\byear{1969}).
\btitle{{Tests for the exponential distribution using Kolmogorov-type
  statistics}}.
\bjournal{J. R. Stat. Soc. Ser. B Stat. Methodol.}
\bvolume{31}
\bpages{499--509}.
\bptok{imsref}%
\end{barticle}
\endbibitem

\bibitem[\protect\citeauthoryear{Teerapabolarn}{2012}]{teera}
\begin{barticle}[mr]
\bauthor{\bsnm{Teerapabolarn},~\bfnm{K.}\binits{K.}}
(\byear{2012}).
\btitle{A pointwise approximation of generalized binomial by {P}oisson
  distribution}.
\bjournal{Appl. Math. Sci. (Ruse)}
\bvolume{6}
\bpages{1095--1104}.
\bid{issn={1312-885X}, mr={2902094}}
\bptok{imsref}%
\end{barticle}\
\endbibitem

\bibitem[\protect\citeauthoryear{Veen and Schoenberg}{2006}]{veenschoenberg}
\begin{bincollection}[author]
\bauthor{\bsnm{Veen},~\bfnm{A.}\binits{A.}} \AND
  \bauthor{\bsnm{Schoenberg},~\bfnm{F.~P.}\binits{F.~P.}}
(\byear{2006}).
\btitle{Assessing spatial point process models using weighted $K$-functions}.
In \bbooktitle{Case Studies in Spatial Point Process Modeling}
(\beditor{\binits{A.}\bfnm{A.}~\bsnm{Baddeley} \betal{et al.}}, eds.).
\bseries{Lecture Notes in Statistics}
\bvolume{185}
\bpages{293--306}.
\bpublisher{Springer}, \blocation{New York}.
\bptok{imsref}%
\end{bincollection}
\endbibitem

\end{thebibliography}
\end{document}